\documentclass[10pt,twoside,twocolumn,english]{IEEEtran}
\usepackage[T1]{fontenc}
\usepackage{color}
\usepackage{array}
\usepackage{float}
\usepackage{units}
\usepackage{multirow}
\usepackage{amsmath}
\usepackage{amssymb}
\usepackage{stackrel}
\usepackage{graphicx}

\makeatletter

\providecommand{\tabularnewline}{\\}
\floatstyle{ruled}
\newfloat{algorithm}{tbp}{loa}
\providecommand{\algorithmname}{Algorithm}
\floatname{algorithm}{\protect\algorithmname}

\ifCLASSOPTIONcompsoc
\usepackage[caption=false,font=normalsize,labelfont=sf,textfont=sf]{subfig}
\else
\usepackage[caption=false,font=footnotesize]{subfig}
\fi
\usepackage{algorithmic}
\usepackage{algorithm}  
\usepackage{cite}
\usepackage{bm}
\usepackage{graphicx}
\usepackage{amsxtra}
\IEEEoverridecommandlockouts

\makeatother

\usepackage{babel}
\begin{document}
\title{\textcolor{black}{Inter-domain Resource Collaboration in Satellite
Networks: An Intelligent Scheduling Approach Towards Hybrid Missions}}
\author{\IEEEauthorblockN{Chenxi Bao, Di Zhou,~\IEEEmembership{Member,~IEEE,}
Min Sheng,~\IEEEmembership{Senior Member,~IEEE,} Yan Shi, Jiandong
Li,~\IEEEmembership{Fellow,~IEEE}}\thanks{Chenxi Bao, Di Zhou, Min Sheng, Yan Shi, and Jiandong Li are with
the State Key Laboratory of Integrated Service Networks, Xidian University,
Xi\textquoteright an, Shaanxi, 710071, China. (e-mail: cxbao@stu.xidian.edu.cn;
\{zhoudi, yshi\}@xidian.edu.cn; \{msheng, jdli\}@mail.xidian.edu.cn).
(Corresponding author: Di Zhou.)}}
\maketitle
\begin{abstract}
\textcolor{black}{Since the next-generation satellite network consisting
of various service function domains, such as communication, observation,
navigation, etc., is moving towards large-scale, using single-domain
resources is difficult to provide satisfied and timely service guarantees
for the rapidly increasing mission demands of each domain. Breaking
the barriers of independence of resources in each domain, and realizing
the cross-domain transmission of missions to efficiently collaborate
inter-domain resources is a promising solution. However, the hybrid
scheduling of different missions and the continuous increase in the
number of service domains have strengthened the differences and dynamics
of mission demands, making it challenging for an efficient cross-domain
mission scheduling (CMS). To this end, this paper first accurately
characterizes the communication resource state of inter-satellite
in real-time exploiting the sparse resource representation scheme,
and systematically characterizes the differentiation of mission demands
by conducting the mission priority model. Based on the information
of resources and missions, we construct the top- and bottom-layer
mission scheduling models of reward association exploiting the correlation
of intra- and inter-domain mission scheduling and formulate the Markov
decision process-based hierarchical CMS problem. Further, to achieve
higher adaptability and autonomy of CMS and efficiently mitigate the
impact of network scale, a hierarchical intelligent CMS algorithm
is developed to dynamically adjust and efficiently match the CMS policy
according to different mission demands. Simulation results demonstrate
that the proposed algorithm has significant performance gain compared
with independent domains and the existing CMS algorithms, and can
still guarantee high service performance under different network scales.}\textcolor{blue}{{}
}\textcolor{black}{}\textcolor{blue}{}
\end{abstract}


\section{Introduction}

\textcolor{black}{The next-generation satellite network is moving
towards multi-functional and large-scale \cite{Deng2020Next,Xie2021LEO,Wang2022Mega,sheng2023coverage},
which includes communication systems that provide diverse service
demands, such as SpaceX and OneWeb \cite{RADTKE2017Interactions,Pultarova2015Telecommunications,Su2019WC},
navigation systems that provide navigation and positioning services,
such as GPS and Beidou, and observation systems that provide various
monitoring and observation services.}\textcolor{black}{{} A service
function system can usually be called a domain. As of now, each domain
is independent of the other, does not share resources, and does not
have data interaction. However, the more significantly differentiated
\cite{Saad2020Network} and rapidly increasing \cite{Ruiz2020Demo}
mission demands make it difficult for using the resources of a single
domain to provide satisfactory and timely service guarantees.}\textcolor{blue}{{}
}\textcolor{black}{ The problem becomes more prominent when encountering
emergencies. }

\textcolor{black}{To cater to the aforementioned issues, breaking
the barriers of independence of resources in each domain, and realizing
the cross-domain transmission of missions to efficiently collaborate
inter-domain resources is considered a promising solution \cite{Hao2022A,He2023Hierarchical}.
However, the attributes of missions in each domain are different.
For example, observation missions (OMs) usually have an enormous data
volume and have low real-time requirements for mission completion.
In contrast, navigation missions (NMs) have a small data volume but
must be completed promptly. Therefore, the cross-domain transmission
of missions will bring missions with multiple attributes for each
domain. The hybrid scheduling of missions with multiple attributes
requires that the scheduling policy can be dynamically adjusted according
to different mission attributes and different mission demands. Besides,
the purpose of the cross-domain transmission is to increase the number
of completed missions of the whole network by inter-domain resource
collaboration when local resources are limited and missions cannot
be offloaded in time. However, the cross-domain transmission should
reduce the impact on the local mission offload of the auxiliary domain.
Therefore, how to design a cross-domain mission scheduling (CMS) scheme
to efficiently collaborate inter-domain resources to maximize the
number of completed missions in the whole network becomes the key
to improving the service performance of satellite networks.}\textcolor{blue}{}

\textcolor{black}{It is technically challenging to develop an efficient
CMS scheme for satellite networks, due to several reasons: 1) the
differentiated mission demands among domains and the coexisted multiple
attributes missions in the domain increase the complexity of effectively
matching mission scheduling policy with demands, which makes effectively
characterizing the mission demands of each domain very important to
achieve efficient CMS; 2) the hybrid scheduling of missions with multiple
attributes strengthens the difference and dynamics of mission demands
and requires higher adaptability and autonomy of CMS policy to satisfy
mission demands; 3) the integration of more satellite systems in satellite
networks has brought about a significant increase in the solving complexity
of CMS, which requires new ideas that are different from traditional
optimization to effectively solve it.}

\textcolor{red}{}\textcolor{black}{In this paper, to address the
technical challenges, we study the CMS problem for satellite networks.
Specifically, we achieve accurate characterization of the communication
resource state of inter-satellite in real-time exploiting the sparse
resource representation (SRR) scheme. Then, we mine the mission characteristics
of each domain to extract mission attributes and model them as the
mission priority to systematically characterize the differentiation
of mission demands. Further, we formulate the CMS problem of maximizing
the number of completed missions satisfying resources and link constraints.
Based on the information of resources and missions and exploiting
the correlation of intra- and inter-domain mission scheduling, we
construct the top- and bottom-layer mission scheduling (TMS/BMS) models
of reward association to convert the CMS problem into the Markov decision
process (MDP)-based hierarchical cross-domain mission scheduling (HCMS)
problem (including TMS and BMS problems) to efficiently solve. Hereafter,
to achieve higher adaptability and autonomy of CMS and efficiently
mitigate the impact of network scale, a hierarchical intelligent CMS
(HICMS) algorithm is developed, which can dynamically adjust and efficiently
match the CMS policy according to different mission demands to achieve
the efficient collaboration of intra- and inter-domain resources to
improve mission completion performance. Extensive simulation results
are provided to demonstrate the performance gains of the HICMS over
existing algorithms.}\textcolor{blue}{}\textcolor{red}{ }\textcolor{black}{}

The key contributions are summarized as follows:
\begin{itemize}
\item \textcolor{black}{The SRR scheme is exploited to achieve the accurate
characterization of the communication resource state to acquire the
connection relationship of inter-satellite (CR-IS) of any two satellites
flexibly and quickly, and the mission priority is modeled to systematically
characterize the differentiation of mission demands, which can quantify
extracted mission attributes.}
\item \textcolor{black}{The CMS problem is converted into the MDP-based
HCMS problem by constructing the TMS and BMS models of reward association
to alleviate the effects brought by the complexity of CMS problems
and the high dynamic of mission demands to efficiently solve.}
\item \textcolor{black}{The HICMS algorithm is developed to solve the HCMS
problem to achieve the efficient collaboration of intra- and inter-domain
resources to improve mission completion performance, which have a
higher adaptability and autonomy of CMS and efficiently mitigate the
impact of network scale.}
\item \textcolor{black}{The extensive simulations are provided to verify
the effectiveness of our proposed scheduling algorithm. From the simulations
we obtain interesting results: 1) CMS brings significant performance
gain compared with independent domains, and HICMS performs better
than the existing CMS algorithms; 2) the performance of HICMS can
still be guaranteed under different network scales.}
\end{itemize}
\setlength{\parindent}{1em}

\textcolor{black}{The rest of this paper is organized as follows.
Section \ref{sec:Related-Work} gives an overview of related works,
followed by a deep model description of the considered satellite network
in Section \ref{sec:System-Model}. Section \ref{sec:Formulation}
formulates the CMS problem and converts it to the MDP-based HCMS problem.
In Section \ref{sec:HICMS-Algorithm-Design}, a HICMS algorithm is
proposed to solve the HCMS problem. The simulation results and discussions
are presented in Section \ref{sec:Simulation-Results-And}, and finally,
conclusions are drawn in Section \ref{sec:Conclusion}.}

\section{\textcolor{black}{Related Work\label{sec:Related-Work}}}

Mission scheduling plays a critical role in efficiently utilizing
the network resources to improve the service performance of networks
\cite{Dai2021Dynamic,Marahatta2021Class,AddZhou2019Distributionally,Qi2020Scalable}.
It can be classified into two main categories, i.e., single-mission
scheduling and hybrid-mission scheduling \cite{zhou2023aerospace}.

\textcolor{red}{}\textcolor{black}{Single-mission scheduling algorithms
focus on scheduling only one type of missions, such as common missions
or burst missions.}\textcolor{red}{{} }Since single-mission scheduling
does no\textcolor{black}{t involve the cross-domain transmission of
missions, the research in this aspect is aimed at the specific domain,
such as the observation domain. Specifically, for common missions,
Di Zhou et al. \cite{Zhou2018Channel} sought the fairness of mission
scheduling to obtain a performance improvement in the number of missions
completed. Lei Wang et al. \cite{Wang2018High} further focused on
users' behavior while mission scheduling and reduced resource conflicts
through user cooperation to improve performance. Furthermore, to utilize
resources and complete missions more efficiently, mission splitting
is considered and significant gains are obtained in \cite{Li2018Graph,Chen2021Task}.
Guohua Wu et al. \cite{Wu2022Flexible} considered the flexibility
of submitted missions and effectively improved the completion rate
of missions by solving mission conflicts. For burst missions, Jianjiang
Wang et al. \cite{Wang2014Dynamic} established the novel multi-objective
dynamic scheduling model for the first time and proposed the mission
merging policy to improve the user satisfaction ratio. To improve
the ability to cope with burst missions, the heuristic and hyperheuristic
algorithms were proposed in \cite{Sun2019JSEE} and \cite{Liu2021A}
respectively.  In a nutshell, the existing single-mission scheduling
algorithm only involves one type of mission, and cannot guarantee
the performance of the hybrid scheduling of missions with multiple
attributes.}

Hybrid-mission scheduling algorithms usually involve \textcolor{black}{common
missions and burst missions}. Specifically, \textcolor{black}{a two-phase
mission scheduling algorithm was proposed and can obtain good performance
in terms of the hybrid missions completion rate \cite{Deng2017Preemptive,Deng2018Two}.
Cui-Qin Dai }\textit{\textcolor{black}{et al}}\textcolor{black}{.
\cite{Dai2021Dynamic} proposed a real-time dynamic scheduling scheme
to guarantee the timely transmission of burst missions and enhance
scheduling efficiency. Although the studies described above involve
hybrid missions and multiple types of user satellites, the relay satellites
assisting mission transmissions are typically geosynchronous and do
not have local missions. Differently, in the cross-domain scenario,
each satellite plays the roles of user satellite and relay satellite,
i.e., no specific relay satellites, and each satellite also assists
others in offloading missions. The cross-domain scheduling of communication
missions (CMs) and OMs was studied in \cite{Hao2022A}, and the proposed
two-stage optimization scheme obtained significant improvements in
scheduling performance. Hongmei He et al. designed the cross-domain
resource scheduling scheme and achieved flexible allocation of missions
\cite{He2023Hierarchical}. However, the available energy resources
for satellites and the differentiation of mission demands in different
domains were not considered in \cite{Hao2022A,He2023Hierarchical}.
This is not practical because the scheduling scheme may need to be
adjusted under the influence of limited energy resources. Further,
in our previous work \cite{Bao2023Towards}, satellite energy resources
and attribute differences have been considered. However, the above
CMS research only focuses on common missions and lacks a systematic
characterization of the mission attributes. }

\textcolor{black}{To sum up, the existing single-mission and hybrid-mission
scheduling algorithms are not adapted to the cross-domain scenario.
Moreover, the existing research on the CMS is very few and needs to
further consider hybrid missions. To this end, this paper investigates
CMS towards hybrid missions and systematically characterize the differentiation
of mission demands. Furthermore, to efficiently collaborate intra-domain
and inter-domain resources, we develop the HICMS algorithm to dynamically
adjust and efficiently match the CMS policy. }

\section{System Model\label{sec:System-Model}}

\subsection{Network Model}

\begin{figure}
\centering{}\includegraphics[width=8cm,height=8cm,keepaspectratio]{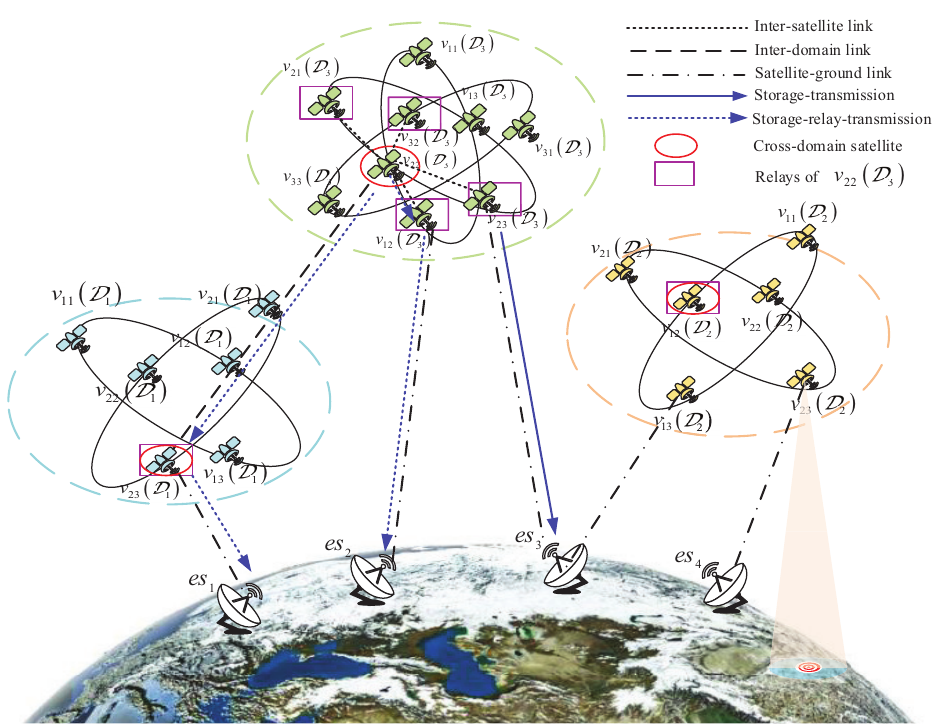}\caption{\label{fig:Satellite-network-scenario}\textcolor{black}{An illustration
of the satellite network consisting of three domains.}}
\end{figure}

\textcolor{black}{We focus on a typical satellite network in this
paper consisting of $K$ domains $\mathcal{D}=\left\{ \mathcal{D}_{1},\cdots,\mathcal{D}_{k},\cdots,\mathcal{D}_{K}\right\} $,
and each domain includes a set of satellites $\mathcal{V}_{k}=\left\{ v_{11}\left(\mathcal{D}_{k}\right),\cdots,v_{ij}\left(\mathcal{D}_{k}\right),\cdots,v_{\mathcal{I}_{k}\mathcal{J}_{k}}\left(\mathcal{D}_{k}\right)\right\} $,
where $v_{ij}\left(\mathcal{D}_{k}\right)$ represents}\textcolor{red}{{}
}\textcolor{black}{the $j$-th satellite in the $i$-th orbit in the
$k$-th domain, $\mathcal{I}_{k}$ is the number of orbit and $\mathcal{J}_{k}$
is the number of satellite in each orbit in $\mathcal{D}_{k}$. In
addition, there are $N$ earth stations represented by $ES=\left\{ es_{1},\cdots,es_{g},\cdots,es_{N}\right\} $.
Figure \ref{fig:Satellite-network-scenario} shows an illustration
of the satellite network consisting of three domains.}\textcolor{blue}{{}
}

\textcolor{black}{In the satellite network, for any domain, the other
domains can be selected as its auxiliary domains}\footnote{\textcolor{black}{Due to limited transceivers of satellites, when
the number of domains is greater than 3, we set that each domain selects
only two domains as auxiliary domains.}}\textcolor{black}{. When satellites complete their missions, they
may also need to assist satellites of other domains to complete mission
transmission}\footnote{\textcolor{black}{In this case, satellites play the role of relay
satellites and are referred to as the \textquotedblleft relay\textquotedblright{}
in this paper.}}\textcolor{black}{. Two transmission modes can be chosen to offload
missions to earth stations: 1) storage-transmission, i.e., satellites
directly transmit missions to earth stations, and 2) storage-relay
transmission, i.e., satellites transmit missions to the relay to offload.
As shown in Fig. \ref{fig:Satellite-network-scenario}, the blue solid
and dotted lines represent the two transmission modes, respectively.
}Furthermore, we make the following assumptions. \textcolor{black}{Firstly,
we divide the planning cycle into $T$ time slots and assume that
the interval of time slot $t$ is fixed as $\tau$, where $t=\left\{ 1,2,\cdots,T\right\} $.
Secondly, we consider the quasi-static network topology. Finally,
we assume that only a transmission link can be established with at
most one earth station for each satellite in each time slot. }\textcolor{blue}{}

\textcolor{black}{We set that each satellite adopts the \textquotedbl one-satellite
four-chain\textquotedbl{} mode to establish inter-satellite links
(ISLs) in the intra-domain, which is widely used in typical satellite
systems \cite{Chen2021Analysis}. Furthermore, we set that satellites
of different domains can transmit missions between domains by establishing
inter-domain links (IDLs). Since not all satellites have IDLs in the
actual network scenario, we select satellites at equal intervals in
each domain to establish IDLs and name these satellites as cross-domain
satellites (CSs). In addition, we set that each CS only establishes
an IDL in each auxiliary domain. In summary, CSs can establish ISLs
and IDLs, and the other satellites (i.e., non-cross-domain satellites
(NCSs)) can only establish ISLs in the intra-domain. In light of the
above rules, the intra-domain relays of each satellite are the satellites
that establish the ISLs with itself and the inter-domain relays of
CSs are the satellites that establish the IDLs with itself.  As shown
in Fig. \ref{fig:Satellite-network-scenario}, taking the satellite
of $\mathcal{D}_{3}$ as an example, $v_{22}\left(\mathcal{D}_{3}\right)$
as the CS is marked with the red oval, and the intra- and inter-domain
relays of $v_{22}\left(\mathcal{D}_{3}\right)$ are marked with the
purple rectangle. $\mathfrak{\mathbb{\mathcal{R}}}\left(v_{ij}\left(\mathcal{D}_{k}\right),\mathcal{D}_{k'}\right)$
is defined as the set of relays, where $k'\in\left\{ 1,\cdots,K\right\} $
is the index of domains. If $k=k'$, $\mathfrak{\mathbb{\mathcal{R}}}\left(v_{ij}\left(\mathcal{D}_{k}\right),\mathcal{D}_{k'}\right)$
is represented as the set of intra-domain relays of $v_{ij}\left(\mathcal{D}_{k}\right)$.
Otherwise, $\mathfrak{\mathbb{\mathcal{R}}}\left(v_{ij}\left(\mathcal{D}_{k}\right),\mathcal{D}_{k'}\right)$
is the set of inter-domain relays of $v_{ij}\left(\mathcal{D}_{k}\right)$
in $D_{k'}$.  }

\vspace{-0.5em}

\subsection{\textcolor{black}{Channel Model}}

\vspace{-0.5em}

\textcolor{black}{We elaborate on the channel model incorporating
the ISL/IDL and satellite-ground link (SGL) in this subsection to
characterize the impact of channel changes on the link transmission
rate. }

\subsubsection{\textcolor{black}{ISL/IDL Channel Model}}

\textcolor{black}{According to recent work \cite{laso2020study,AddZhou2019Collaborative},
the achievable transmission rate of the link from $v_{ij}\left(\mathcal{D}_{k}\right)$
to its relay in the $t$-th time slot, denoted by $Cs^{t}\left({\scriptstyle v_{ij}\left(\mathcal{D}_{k}\right),m}\right)$,
can be expressed as:}

\vspace{-1.0em}\textcolor{black}{
\begin{equation}
Cs^{t}\left({\scriptstyle v_{ij}\left(\mathcal{D}_{k}\right),m}\right)=\frac{P_{sst}G_{tr}\left(v_{ij}\left(\mathcal{D}_{k}\right)\right)G_{re}\left(m\right)}{\mathcal{K}T_{n}\cdot\left(\nicefrac{E_{b}}{N_{0}}\right)_{req}\cdot\mathcal{L}^{t}\left(v_{ij}\left(\mathcal{D}_{k}\right),m\right)\cdot\mathcal{\textrm{M}}},
\end{equation}
where $m\in\mathfrak{\mathbb{\mathcal{R}}}\left(v_{ij}\left(\mathcal{D}_{k}\right),\mathcal{D}_{k'}\right)$
is the relay and $\mathcal{L}^{t}\left(v_{ij}\left(\mathcal{D}_{k}\right),m\right)$
is the free space loss shown as follows:
\begin{equation}
\mathcal{L}^{t}\left(v_{ij}\left(\mathcal{D}_{k}\right),m\right)=\left(\frac{4\pi\cdot Sl^{t}\left(v_{ij}\left(\mathcal{D}_{k}\right),m\right)\cdot\mathcal{F}}{c}\right)^{2}.
\end{equation}
$Sl^{t}\left(v_{ij}\left(\mathcal{D}_{k}\right),m\right)$ is the
slant range between $v_{ij}\left(\mathcal{D}_{k}\right)$ and its
relay. $\mathcal{F}$ and $c$ are the communication center frequency
and speed of light. $P_{sst}$ is the constant transmission power
of satellites for ISL/IDL. $G_{tr}\left(v_{ij}\left(\mathcal{D}_{k}\right)\right)$
and $G_{re}\left(m\right)$ are the transmitting and receiving antenna
gain respectively. Besides, $\mathcal{K}$ is the Boltzmann\textquoteright s
constant, and $T_{n}$ is the total system noise temperature. $\left(\nicefrac{E_{b}}{N_{0}}\right)_{req}$
and $\textrm{M}$ are the required ratio of received energy-per-bit
to noise-density, and link margin.}

\subsubsection{\textcolor{black}{SGL Channel Model}}

\textcolor{black}{The achievable transmission rate of the link from
$v_{ij}\left(\mathcal{D}_{k}\right)$ to $es_{g}$ in the $t$-th
time slot is calculated using the Shannon formula \cite{Fu2020Integrated},
denoted by $Ce^{t}\left(v_{ij}\left(\mathcal{D}_{k}\right),es_{g}\right)$,}

\textcolor{black}{
\begin{equation}
Ce^{t}\!\left(v_{ij}\left(\mathcal{D}_{k}\right)\!,\!es_{g}\right)\!=\!\mathcal{B}\!\cdot\!\textrm{log}_{2}\left(\!1\!+\!\mathcal{SNR}^{t}\!\left(\!v_{ij}\left(\mathcal{D}_{k}\right)\!,\!es_{g}\!\right)\right)\!,
\end{equation}
where $\mathcal{SNR}^{t}\left(v_{ij}\left(\mathcal{D}_{k}\right),es_{g}\right)$
is the signal-to-noise ratio, expressed as:}

\textcolor{black}{
\begin{equation}
\mathcal{SNR}^{t}\left(v_{ij}\left(\mathcal{D}_{k}\right),es_{g}\right)=\frac{P_{set}G_{tr}\left(v_{ij}\left(\mathcal{D}_{k}\right)\right)G_{re}\left(es_{g}\right)\mathcal{L}_{p}^{t}}{\mathcal{N}\cdot\mathcal{L}^{t}\left(v_{ij}\left(\mathcal{D}_{k}\right),es_{g}\right)}.
\end{equation}
$P_{set}$ is the constant transmission power of satellites for SGL.
$\mathcal{L}_{p}^{t}$ is the propagation loss. $\mathcal{N}$ is
the noise. $\mathcal{B}$ is the available transmission bandwidth.}\textcolor{blue}{}\textcolor{black}{{}
}

\subsection{\textcolor{black}{Sparse Resource  Representation}}

\textcolor{black}{Due to the high-speed orbiting movement of satellites,
the inter-satellite communication resources are highly dynamic \cite{AddLiu2016An}.
We use the SRR scheme to characterize CR-IS of ISL/IDL, denoted by
$l^{t}\left(v_{ij}\left(\mathcal{D}_{k}\right),m\right)$, where $m\in\mathfrak{\mathbb{\mathcal{R}}}\left(v_{ij}\left(\mathcal{D}_{k}\right)\right)$
is the relay and $\mathfrak{\mathbb{\mathcal{R}}}\left(v_{ij}\left(\mathcal{D}_{k}\right)\right)$
$=\left\{ \mathfrak{\mathbb{\mathcal{R}}}\left(v_{ij}\left(\mathcal{D}_{k}\right),\mathcal{D}_{k'}\right)\left|k'\!\in\!\left\{ 1,2,\cdots,K\right\} \right.\right\} $
is the set of all relays of $v_{ij}\left(\mathcal{D}_{k}\right)$
\cite{Bao2023Towards}.}\textcolor{blue}{{} }

\textcolor{black}{The SRR scheme characterizes the CR-IS of ISL/IDL
in real-time by constructing the satellite orbit feature matrix (SOFM)
of each domain to deduce the network topology. Specifically, the SOFM
of each domain consists of orbit parameters representing the satellite
position of the initial time slot, including the orbit height $h_{i}$,
eccentricity $e_{i}$, inclination $I_{i}$, argument of perigee $\Omega_{i}$,
right ascension of ascending node $RA_{i}$, and mean anomaly in the
first time slot $\overline{TA_{ij}^{1}}$, where $i\in\left\{ 1,2,\cdots,\mathcal{I}_{k}\right\} ,j\in\left\{ 1,2,\cdots,\mathcal{J}_{k}\right\} $.
Using the position information of the satellites obtained by searching
the SOFM, the maximum geocentric angle $\theta^{max}\left(v_{ij}\left(\mathcal{D}_{k}\right)\text{,\ensuremath{v_{ij}\left(\mathcal{D}_{k\text{'}}\right)}}\right)$
and geocentric angle $\theta^{t}\left(v_{ij}\left(\mathcal{D}_{k}\right)\text{,\ensuremath{v_{ij}\left(\mathcal{D}_{k\text{'}}\right)}}\right)$
in the $t$-th time slot can be calculated to acquire the CR-IS of
ISL/IDL. If $\theta^{t}\left(v_{ij}\left(\mathcal{D}_{k}\right)\text{,\ensuremath{v_{ij}\left(\mathcal{D}_{k\text{'}}\right)}}\right)\leq\theta^{max}\left(v_{ij}\left(\mathcal{D}_{k}\right)\text{,\ensuremath{v_{ij}\left(\mathcal{D}_{k\text{'}}\right)}}\right)$,
the ISL/IDL can be established, i.e., $l^{t}\left(v_{ij}\left(\mathcal{D}_{k}\right),m\right)=1$.
Otherwise, $l^{t}\left(v_{ij}\left(\mathcal{D}_{k}\right),m\right)=0$.
}\textcolor{black}{\footnotesize{}}\textcolor{black}{}

\subsection{Mission Priority Model}

\textcolor{black}{We define that the mission of each domain consists
of a 5-tuple. Specifically, $\mathcal{M}^{t}\left(v_{ij}\left(\mathcal{D}_{k}\right)\right)$=$\left\{ \mathcal{P}\left(\mathcal{M}_{k}\right),\rho\left(\mathcal{M}_{k}\right),t,\mathcal{R}s\left(\mathcal{M}_{k}\right),v_{ij}\left(\mathcal{D}_{k}\right)\right\} $
represents the mission collected/generated by $v_{ij}\left(\mathcal{D}_{k}\right)$
in the $t$-th time slot, where $\mathcal{M}_{k}$ represents the
mission of $\mathcal{D}_{k}$. $\mathcal{P}\left(\mathcal{M}_{k}\right)$
represents the priority of $\mathcal{M}_{k}$, which is used to determine
the order in which each mission is served in different domains. $\rho\left(\mathcal{M}_{k}\right)$
represents the data volume of $\mathcal{M}_{k}$. $t$ represents
the mission collected/generated in the $t$-th time slot. $\mathcal{R}s\left(\mathcal{M}_{k}\right)$
represents the survival time slots of $\mathcal{M}_{k}$. Furthermore,
$\mathcal{R}s\left(\mathcal{M}_{k}\right)$ is decremented with a
step size of 1 during the CMS, i.e., the remaining survival time slot
of each mission is different in each time slot, and when decremented
to 0, the mission will be deleted from the on-board memory, i.e.,
the mission is not completed. $v_{ij}\left(\mathcal{D}_{k}\right)$
is used to clarify the reward allocation during training. In this
paper, we set that there are common missions and burst missions in
each domain and do not consider the splitting of missions, i.e., each
mission must complete the transmission on one path.}\textcolor{red}{{}
}

\textcolor{black}{We extract mission attributes from three aspects
(i.e., the data volume of the mission, mission arrival rate, and delay
tolerance) and model them as the mission priority to effectively characterize
the differentiation of mission demands}\footnote{\textcolor{black}{It should be noted that this paper only establishes
the priority model of common missions, and the priority of burst missions
is set higher than that of all common missions.}}\textcolor{black}{. The calculation formulation of mission priority
is as follows \cite{Wu2019Research}: }

\textcolor{blue}{\vspace{-0.5em}}\textcolor{black}{
\begin{equation}
\mathcal{P}\left(\mathcal{M}_{k}\right)=\stackrel[{\scriptstyle f=1}]{{\scriptstyle \mathcal{F}}}{\sum}\mathcal{W}_{f}\left(\mathcal{M}_{k}\right)\cdot\mathcal{V}_{f}\left(\mathcal{M}_{k}\right),\label{eq:priority}
\end{equation}
where $\mathcal{F}=3$ is the number of the mission attributes. $\mathcal{W}_{f}\left(\mathcal{M}_{k}\right)$
represents the weight of the $f$-th mission attribute of $\mathcal{M}_{k}$,
which examines the degree of influence of each mission attribute on
the priority of the mission. $\mathcal{W}_{f}\left(\mathcal{M}_{k}\right)$
is shown as follows:}

\textcolor{blue}{\vspace{-0.5em}}\textcolor{black}{
\begin{equation}
\mathcal{W}_{f}\left(\mathcal{M}_{k}\right)=\frac{\stackrel[y=1]{\mathcal{F}}{\sum}\mathcal{\textrm{I}}_{fy}\left(\mathcal{M}_{k}\right)}{\stackrel[x=1]{\mathcal{F}}{\sum}\stackrel[y=1]{\mathcal{F}}{\sum}\mathcal{\textrm{I}}_{xy}\left(\mathcal{M}_{k}\right)},\:\left\{ f=1,\cdots,\mathcal{F}\right\} ,
\end{equation}
where $\mathcal{\textrm{I}}_{xy}\left(\mathcal{M}_{k}\right)$ is
a score obtained by comparing the importance of each mission attribute
using a 0-1 scoring method to effectively evaluate the weight of each
mission attribute in $\mathcal{M}_{k}$. $\mathcal{\textrm{I}}_{xy}\left(\mathcal{M}_{k}\right)$
is expressed as follows:}

\textcolor{blue}{\vspace{-0.5em}}\textcolor{black}{
\begin{equation}
\mathcal{\textrm{I}}_{xy}\left(\mathcal{M}_{k}\right)=\begin{cases}
1 & x>y,\\
0.5 & x=y,\\
0 & x<y.
\end{cases}
\end{equation}
$\mathcal{V}_{f}\left(\mathcal{M}_{k}\right)$ represents the quantitative
value of the $f$-th mission attribute of $\mathcal{M}_{k}$, and
the specific value reflects the relationship of the $f$-th mission
attribute among different domain missions.}\textcolor{red}{{} }\textcolor{black}{For
example, the NM should be offloaded in a timely to guarantee its effectiveness,
and the OM has a higher delay tolerance than the NM. Therefore, the
quantitative value of the delay tolerance attribute of the NM is set
higher than that of OM.}

\subsection{Energy Model}

\textcolor{black}{In this subsection, we introduce the energy consumption
model and the energy harvesting model, respectively. To model the
energy consumption of satellites, we define $\mathcal{H}_{\mathcal{M}_{k'}}^{t}\left(v_{ij}\left(\mathcal{D}_{k}\right),m\right)$
as the number of $\mathcal{M}_{k'}$ that $v_{ij}\left(\mathcal{D}_{k}\right)$
transmits to its relay}\textcolor{blue}{}\footnote{\textcolor{black}{Since we consider the CMS for satellite networks,
CSs can receive the missions of any domain by ISLs and IDLs, and NCSs
can also receive missions of the other domains by establishing ISL
with CSs.}}\textcolor{black}{{} and $\mathcal{H}_{\mathcal{M}_{k'}}^{t}\left(v_{ij}\left(\mathcal{D}_{k}\right),es_{g}\right)$
as the number of $\mathcal{M}_{k'}$ that $v_{ij}\left(\mathcal{D}_{k}\right)$
transmits to $es_{g}$ during the $t$-th time slot. Hereafter, we
elaborate on the two models.}\textcolor{blue}{{} }\textcolor{black}{}

\textcolor{black}{The energy consumption of $v_{ij}\left(\mathcal{D}_{k}\right)$
for transmitting data during the $t$-th time slot, is expressed as:}

\textcolor{black}{
\begin{equation}
\begin{array}{cc}
E_{tr}^{t}\left(v_{ij}\left(\mathcal{D}_{k}\right)\right)\!\!= & \!\!\!\!\underset{{\scriptstyle m\in\mathfrak{\mathbb{\mathcal{R}}}\left(v_{ij}\left(\mathcal{D}_{k}\right)\right)}}{\sum}\!\!\frac{\mathcal{X}^{t}\left(v_{ij}\left(\mathcal{D}_{k}\right),m\right)}{Cs^{t}\left(v_{ij}\left(\mathcal{D}_{k}\right),m\right)}\!\cdot\!P_{sst}\\
 & +\stackrel[{\scriptstyle g=1}]{{\scriptstyle N}}{\sum}\frac{{\scriptstyle \mathcal{X}^{t}\left(v_{ij}\left(\mathcal{D}_{k}\right),es_{g}\right)}}{{\scriptstyle Ce^{t}\left(v_{ij}\left(\mathcal{D}_{k}\right),es_{g}\right)}}\cdot P_{set},
\end{array}
\end{equation}
where $\mathcal{X}^{t}\left(v_{ij}\left(\mathcal{D}_{k}\right),m\right)=\stackrel[k'=1]{K}{\sum}\mathcal{H}_{\mathcal{M}_{k'}}^{t}\left(v_{ij}\left(\mathcal{D}_{k}\right),m\right)\cdot\rho\left(\mathcal{M}_{k'}\right)$
represents the data volume that $v_{ij}\left(\mathcal{D}_{k}\right)$
transmits to its relay, and $\mathcal{X}^{t}\left(v_{ij}\left(\mathcal{D}_{k}\right),es_{g}\right)=\stackrel[k'=1]{K}{\sum}\mathcal{H}_{\mathcal{M}_{k'}}^{t}\left(v_{ij}\left(\mathcal{D}_{k}\right),es_{g}\right)\cdot\rho\left(\mathcal{M}_{k'}\right)$
represents the data volume that $v_{ij}\left(\mathcal{D}_{k}\right)$
transmits to $es_{g}$. }

\textcolor{black}{The energy consumption of $v_{ij}\left(\mathcal{D}_{k}\right)$
for receiving data is denoted as $E_{r}^{t}\left(v_{ij}\left(\mathcal{D}_{k}\right)\right)$
during the $t$-th time slot, as follows:
\begin{equation}
E_{r}^{t}\left(v_{ij}\left(\mathcal{D}_{k}\right)\right)\!\!=\underset{{\scriptstyle m\in\mathbb{R}\left(v_{ij}\left(\mathcal{D}_{k}\right)\right)}}{\sum}\!\!\frac{\widehat{\mathcal{X}^{t}}\left(m,v_{ij}\left(\mathcal{D}_{k}\right)\right)}{Cs^{t}\left(m,v_{ij}\left(\mathcal{D}_{k}\right)\right)}\!\cdot\!P_{sr},
\end{equation}
where $\mathbb{R}\left(v_{ij}\left(\mathcal{D}_{k}\right)\right)\!=\!\left\{ \!v_{i'j'}\!\left(\mathcal{D}_{k'}\right)\!\!\right.\left|\!v_{ij}\left(\mathcal{D}_{k}\right)\!\in\!\mathfrak{\mathcal{R}\!}\left(\!v_{i'j'}\left(\mathcal{D}_{k'}\right)\!,\!\mathcal{D}_{k}\right)\!,\right.$
$\left.\left.i'\in\left\{ 1,2,\cdots,\mathcal{I}_{k'}\right\} \!,j'\in\left\{ 1,2,\right.\cdots,\mathcal{J}_{k'}\right\} ,k'\!\in\!\left\{ 1,2,\cdots,K\right\} \right\} $
is the set of satellites that can transmit missions to $v_{ij}\left(\mathcal{D}_{k}\right)$.
$\widehat{\mathcal{X}^{t}}\left(m,v_{ij}\left(\mathcal{D}_{k}\right)\right)$
represents the data volume received by $v_{ij}\left(\mathcal{D}_{k}\right)$
originating from $m$. It should be noted that due to the limitation
of storage capacity and remaining battery energy, $\widehat{\mathcal{X}^{t}}\left(m,v_{ij}\left(\mathcal{D}_{k}\right)\right)\leq\mathcal{X}^{t}\left(m,v_{ij}\left(\mathcal{D}_{k}\right)\right)$.
$P_{sr}$ is the reception power of satellites.}

\textcolor{black}{Moreover, we define the energy consumption of $v_{ij}\left(\mathcal{D}_{k}\right)$
for nominal operation during the $t$-th time slot, denoted by $E_{o}^{t}\left(v_{ij}\left(\mathcal{D}_{k}\right)\right)$,
is expressed as follows:}

\textcolor{blue}{\vspace{-0.5em}}\textcolor{black}{
\begin{equation}
E_{o}^{t}\left(v_{ij}\left(\mathcal{D}_{k}\right)\right)=P_{o}\cdot\tau,
\end{equation}
where $P_{o}$ is the nominal operation power.}

\textcolor{black}{The total energy consumption of $v_{ij}\left(\mathcal{D}_{k}\right)$
during the $t$-th time slot consists of the above energy consumption
items, denoted by $E_{c}^{t}\left(v_{ij}\left(\mathcal{D}_{k}\right)\right)$,
as follows:}

\textcolor{blue}{\vspace{-0.5em}}\textcolor{black}{
\begin{equation}
E_{c}^{t}\!\left(v_{ij}\left(\mathcal{D}_{k}\right)\right)\!\!=\!\!E_{tr}^{t}\!\left(\!v_{ij}\left(\mathcal{D}_{k}\right)\!\right)\!+\!E_{r}^{t}\!\left(\!v_{ij}\left(\mathcal{D}_{k}\right)\!\right)\!+\!E_{o}^{t}\!\left(\!v_{ij}\left(\mathcal{D}_{k}\right)\!\right)\!.
\end{equation}
}

\textcolor{black}{We denote $E_{h}^{t}\left(v_{ij}\left(\mathcal{D}_{k}\right)\right)$
as the harvested energy of $v_{ij}\left(\mathcal{D}_{k}\right)$ during
the $t$-th time slot and determine it in advance according to orbital
dynamics. $E_{h}^{t}\left(v_{ij}\left(\mathcal{D}_{k}\right)\right)$
is expressed as:
\begin{equation}
E_{h}^{t}\left(v_{ij}\left(\mathcal{D}_{k}\right)\right)=P_{h}\cdot\min\left\{ \tau,y^{t}\left(v_{ij}\left(\mathcal{D}_{k}\right)\right)\right\} ,
\end{equation}
where $P_{h}$ is the energy collection rate. $y^{t}\left(v_{ij}\left(\mathcal{D}_{k}\right)\right)$
is the duration that the satellite is covered by the sun at the beginning
of the $t$-th time slot. When the satellite is completely eclipsed
by the Earth in the $t$-th time slot, $y^{t}\left(v_{ij}\left(\mathcal{D}_{k}\right)\right)=0$.}

\textcolor{blue}{\vspace{-1.0em}}

\section{F\textcolor{black}{ormulation\label{sec:Formulation}}}

\textcolor{black}{In this section, we study the CMS problem of maximizing
the number of completed missions satisfying the resources and link
constraints. First of all, we present the related constraints, and
then the proposed CMS problem is formulated based on these constraints.
Finally, we convert it into the MDP-based HCMS problem to solve.}\textcolor{red}{}

\textcolor{blue}{\vspace{-1.0em}}

\subsection{Related Constraints Modeling}

\textcolor{blue}{\vspace{-0.5em}}

\subsubsection{Storage Resource Constraint}

\textcolor{black}{The total data volume of all missions stored on
each satellite cannot exceed its storage capacity,
\begin{equation}
\mathcal{B}^{t}\left(v_{ij}\left(\mathcal{D}_{k}\right)\right)\!+\widehat{\mathcal{X}^{t}}\left(v_{ij}\left(\mathcal{D}_{k}\right)\right)\!-\mathcal{X}^{t}\left(v_{ij}\left(\mathcal{D}_{k}\right)\right)\!+\!\rho\left(\mathcal{M}_{k}\right)\!\leq\!\mathcal{B}_{max}.\label{eq:Storage}
\end{equation}
where $\mathcal{B}^{t}\left(v_{ij}\left(\mathcal{D}_{k}\right)\right)$
is the total data volume of missions stored in $v_{ij}\left(\mathcal{D}_{k}\right)$
at the beginning of $t$-th time slot, $\widehat{\mathcal{X}^{t}}\left(v_{ij}\left(\mathcal{D}_{k}\right)\right)=\underset{{\scriptstyle m\in\mathbb{R}\left(v_{ij}\left(\mathcal{D}_{k}\right)\right)}}{\sum}\!\!\widehat{\mathcal{X}^{t}}\left(m,v_{ij}\left(\mathcal{D}_{k}\right)\right)$
is the total data volume received by $v_{ij}\left(D_{k}\right)$,
$\mathcal{X}^{t}\left(v_{ij}\left(\mathcal{D}_{k}\right)\right)=\!\!\underset{{\scriptstyle {\scriptstyle m\in\mathfrak{\mathbb{\mathcal{R}}}\left(v_{ij}\left(\mathcal{D}_{k}\right)\right)}}}{\sum}\!\!\mathcal{X}^{t}\left(v_{ij}\left(\mathcal{D}_{k}\right),m\right)+\!\!\!\stackrel[{\scriptstyle g=1}]{{\scriptstyle N}}{\sum}\!\!\mathcal{X}^{t}\left(v_{ij}\left(\mathcal{D}_{k}\right),es_{g}\right)$
is the total data volume that can be transmitted by $v_{ij}\left(\mathcal{D}_{k}\right)$,
and $\mathcal{B}_{max}$ is the storage capacity}\footnote{\textcolor{black}{It should be noted that each domain includes CSs
and NCSs, and CSs need to undertake the intra-domain and cross-domain
mission's transmission and auxiliary work. Therefore, this paper sets
that CSs have more resources than NCSs, i.e., CSs have a higher storage
and battery capacity.}}\textcolor{black}{. }

\subsubsection{\textcolor{black}{Energy Resource Constraint}}

\textcolor{black}{The battery capacity of each satellite is also limited,
and the normal operation of the satellite in shadow should be maintained.
Therefore, all energy cannot be fully used for the mission transmission
and reception, i.e., 
\begin{equation}
E^{t}\left(v_{ij}\left(\mathcal{D}_{k}\right)\right)-E_{c}^{t}\left(v_{ij}\left(\mathcal{D}_{k}\right)\right)\geq E_{min},
\end{equation}
and the battery energy cannot exceed the battery capacity, i.e.,
\begin{equation}
E^{t}\left(v_{ij}\left(\mathcal{D}_{k}\right)\right)+\widetilde{E_{h}^{t}}\left(v_{ij}\left(\mathcal{D}_{k}\right)\right)-E_{c}^{t}\left(v_{ij}\left(\mathcal{D}_{k}\right)\right)\leq E_{max},
\end{equation}
where $E^{t}\left(v_{ij}\left(\mathcal{D}_{k}\right)\right)$ is the
residual energy in $v_{ij}\left(\mathcal{D}_{k}\right)$ at the beginning
of the $t$-th time slot, $E_{min}=E_{max}-\eta\cdot E_{max}$ represents
the minimum residual energy of the battery and $\eta$ represents
the maximum discharge depth of the battery. $\widetilde{E_{h}^{t}}\left(v_{ij}\left(\mathcal{D}_{k}\right)\right)\leq E_{h}^{t}\left(v_{ij}\left(\mathcal{D}_{k}\right)\right)$
is the actual harvested energy, and $E_{max}$ is the battery capacity.}

\subsubsection{\textcolor{black}{Communication Resource Constraints}}

\textcolor{black}{During the mission transmission, since the transmission
rate of the link is limited, each satellite transmits the total data
volume that cannot exceed the link capacity. Furthermore, we consider
that each satellite only transmits missions to one relay or one earth
station in each time slot. Therefore, binary variables $u^{t}\left(v_{ij}\left(\mathcal{D}_{k}\right),m\right)$
and $u^{t}\left(v_{ij}\left(\mathcal{D}_{k}\right),es_{g}\right)$
$\in\left\{ 0,1\right\} $ are introduced to represent whether the
links from $v_{ij}\left(\mathcal{D}_{k}\right)$ to the relays/earth
stations are used, 1 if used and 0 otherwise. The communication resource
constraints are as follows:}

\textcolor{black}{
\begin{equation}
\begin{array}{cc}
\mathcal{X}^{t}\!\left(\!v_{ij}\left(\mathcal{D}_{k}\right)\!\right)\!\leq & \!\!\!\!\!\left(\!\!\underset{{\scriptstyle {\scriptstyle m\in\mathfrak{\mathbb{\mathcal{R}}}\left(v_{ij}\left(\mathcal{D}_{k}\right)\right)}}}{\sum}\!\!\!\!\!\!\!Cs^{t}\left(v_{ij}\left(\mathcal{D}_{k}\right),m\right)\!\cdot\!u^{t}\left(v_{ij}\!\left(\mathcal{D}_{k}\right)\!,\!m\!\right)\right.\\
 & \left.\!\!\!\!\!\!\!\!\!\!\!\!\!\!\!+\!\!\stackrel[{\scriptstyle g=1}]{{\scriptstyle N}}{\sum}Ce^{t}\left(v_{ij}\left(\mathcal{D}_{k}\right),es_{g}\right)\!\cdot\!u^{t}\left(\!v_{ij}\left(\mathcal{D}_{k}\right)\!,\!es_{g}\right)\!\right)\!\cdot\!\tau
\end{array}
\end{equation}
}

\subsubsection{\textcolor{black}{Link Constraint}}

\textcolor{black}{Since we consider that each satellite only transmits
missions to one relay or one earth station in each time slot, only
one link can be selected for the mission transmission of each satellite
in each time slot,}

\textcolor{black}{
\begin{equation}
\underset{{\scriptstyle m\in\mathfrak{\mathbb{\mathcal{R}}}\left(v_{ij}\left(\mathcal{D}_{k}\right)\right)}}{\sum}\!\!\!\!\!\!\!\!\!u^{t}\left(v_{ij}\left(\mathcal{D}_{k}\right),m\right)\!\!+\!\!\stackrel[{\scriptstyle g=1}]{{\scriptstyle N}}{\sum}\!u^{t}\!\left(\!v_{ij}\left(\mathcal{D}_{k}\right)\!,\!es_{g}\right)\!\!=\!\!1,\label{eq:Link}
\end{equation}
}

\vspace{-0.5em}

\subsection{\textcolor{black}{Problem Formulation}}

\textcolor{black}{To achieve effective offloading of missions in
each domain, we jointly consider the resources and link constraints
to maximize the number of completed missions in the whole network.
Mathematically, this problem can be formulated as follows:
\begin{equation}
\begin{array}{c}
\!\!\negthickspace\negthickspace\textrm{CMS:\!\!}\underset{\underset{{\scriptstyle u^{t}\left(\!v_{ij}\left(\mathcal{D}_{k}\right),es_{g}\!\right)}}{u^{t}\left(\!v_{ij}\left(\mathcal{D}_{k}\right),m\!\right)}}{\max}\underset{k\in\left\{ 1,2,\cdots,K\right\} }{\sum}\mathbb{C}\left(\mathcal{D}_{k}\right)\\
\textrm{s.t.}\enskip\left(\ref{eq:Storage}\right)-\left(\ref{eq:Link}\right),
\end{array}
\end{equation}
where $\mathcal{\mathbb{C}}\left(\mathcal{D}_{k}\right)=\stackrel[{\scriptstyle i=1}]{{\scriptscriptstyle \mathcal{{\scriptstyle I}}_{k}}}{\sum}\stackrel[{\scriptstyle j=1}]{{\scriptscriptstyle \mathcal{{\scriptstyle J}}_{k}}}{\sum}\stackrel[{\scriptstyle t=1}]{{\scriptstyle T}}{\sum}\stackrel[{\scriptstyle g=1}]{{\scriptstyle N}}{\sum}\mathcal{H}^{t}\left(v_{ij}\left(\mathcal{D}_{k}\right),es_{g}\right)$
is the total number of completed missions in $\mathcal{D}_{k}$, and
$\mathcal{H}^{t}\left(v_{ij}\left(\mathcal{D}_{k}\right),es_{g}\right)=\stackrel[k'=1]{K}{\sum}\mathcal{H}_{\mathcal{M}_{k'}}^{t}\left(v_{ij}\left(\mathcal{D}_{k}\right),es_{g}\right).$
It should be noted that although this formulation only includes missions
that are offloaded on the SGL, the missions may be offloaded by mode
(2), i.e., storage-relay-transmission.}

\textcolor{black}{In this formulation, decision variables are integer
variables. Therefore, the CMS problem is the integer linear programming
problem, which is NP-hard and mission scheduling demands are dynamic,
which strengthens the solution complexity of the problem. Furthermore,
with the continuous increase of satellite systems and the continuous
expansion of the constellations' scale, the number of decision variables
for the CMS problem increases rapidly, which makes it more difficult
to solve the CMS problem using the traditional optimization method.
Therefore, we need to convert the CMS problem so that it can be solved
efficiently.}

\vspace{-0.5em}

\subsection{CMS Problem Conversion}

\textcolor{black}{In this subsection, exploiting the Markov property
of the CMS process and the correlation of intra- and inter-domain
mission scheduling, we convert the CMS problem into the MDP-based
HCMS problem (including TMS and BMS problems) by constructing the
TMS and BMS models of reward association. Specifically, the BMS problem
is formulated to solve the mission scheduling in the domain to achieve
intra-domain resource collaboration, and the TMS problem is formulated
to solve the mission scheduling in the inter-domain to achieve inter-domain
resource collaboration.}\textcolor{red}{}\textcolor{black}{}

\subsubsection{\textcolor{black}{BMS Problem}}

\textcolor{black}{Before describing the BMS problem, we first define
the state information of each satellite. Due to the obvious differences
in the resources of CSs and NCSs and each domain mission attribute,
we adopt the relative value of the resource states and the mission
states to replace the true value to describe each satellite. Specifically,
we set that the state of each satellite $s^{t}\left(v_{ij}\left(\mathcal{D}_{k}\right)\right)$
consists of a 4-tuple, i.e., 
\begin{equation}
\begin{array}{cc}
s^{t}\left(v_{ij}\left(\mathcal{D}_{k}\right)\right)= & \!\!\!\!\!\!\left(\widetilde{\mathcal{B}^{t}}\left(v_{ij}\left(\mathcal{D}_{k}\right)\right),\widetilde{E^{t}}\left(v_{ij}\left(\mathcal{D}_{k}\right)\right)\text{,}\right.\\
 & \left.\widetilde{Ce^{t}}\left(v_{ij}\left(\mathcal{D}_{k}\right)\right),\widetilde{\mathcal{R}s^{t}}\left(v_{ij}\left(\mathcal{D}_{k}\right)\right)\right),
\end{array}
\end{equation}
where $\widetilde{\mathcal{B}^{t}}\left(v_{ij}\left(\mathcal{D}_{k}\right)\right)$
is the relative value that characterizes storage occupied, $\widetilde{E^{t}}\left(v_{ij}\left(\mathcal{D}_{k}\right)\right)$
is the relative value that characterizes remaining available battery
energy, $\widetilde{Ce^{t}}\left(v_{ij}\left(\mathcal{D}_{k}\right)\right)$
indicates whether can directly offload the missions to earth stations,
and $\widetilde{\mathcal{R}s^{t}}\left(v_{ij}\left(\mathcal{D}_{k}\right)\right)$
is the average of the remaining survival time slots for all missions
stored on $v_{ij}\left(\mathcal{D}_{k}\right)$ at the beginning of
the $t$-th time slot.}

\textcolor{black}{Specifically, $\widetilde{\mathcal{B}^{t}}\left(v_{ij}\left(\mathcal{D}_{k}\right)\right)=\nicefrac{\mathcal{B}_{max}}{\mathcal{B}^{t}\left(v_{ij}\left(\mathcal{D}_{k}\right)\right)}$
and if $\mathcal{B}^{t}\left(v_{ij}\left(\mathcal{D}_{k}\right)\right)=0$,
$\widetilde{\mathcal{B}^{t}}\left(v_{ij}\left(\mathcal{D}_{k}\right)\right)=\mathcal{B}_{max}$.
$\mathcal{B}^{t}\left(v_{ij}\left(\mathcal{D}_{k}\right)\right)\in\left[0,\mathcal{B}_{max}\right]$
is denoted as}

\textcolor{black}{
\begin{equation}
\!\begin{array}{cc}
\mathcal{B}^{t}\left(v_{ij}\left(\mathcal{D}_{k}\right)\right)\!= & \mathcal{B}^{t-1}\left(v_{ij}\left(\mathcal{D}_{k}\right)\right)\!+\!\widehat{\mathcal{X}^{t-1}}\left(v_{ij}\left(\mathcal{D}_{k}\right)\right)\!\\
 & -\!\mathcal{X}^{t-1}\left(v_{ij}\left(\mathcal{D}_{k}\right)\right)\!+\!\rho\left(\mathcal{M}_{k}\right)\!.
\end{array}\label{eq:Bt}
\end{equation}
}\vspace{-0.5em}

\textcolor{black}{$\widetilde{E^{t}}\left(\!v_{ij}\left(\mathcal{D}_{k}\right)\!\right)\!=\!\nicefrac{E^{t}\left(v_{ij}\left(\mathcal{D}_{k}\right)\right)}{\left(E_{min}+E_{o}^{t}\left(v_{ij}\left(\mathcal{D}_{k}\right)\right)\right)}$.}\textcolor{blue}{{}
}\textcolor{black}{$E^{t}\left(\!v_{ij}\left(\mathcal{D}_{k}\right)\!\right)$
$\in\left[0,E_{max}\right]$ is denoted as
\begin{equation}
E^{t}\!\!\left({\scriptstyle v_{ij}\left(\mathcal{D}_{k}\right)}\right)\!=\!E^{t-1}\!\left({\scriptstyle v_{ij}\left(\mathcal{D}_{k}\right)}\right)\!+\!\widetilde{E_{h}^{t-1}\!}\left({\scriptstyle v_{ij}\left(\mathcal{D}_{k}\right)}\right)\!-\!E_{c}^{t-1}\!\left({\scriptstyle v_{ij}\left(\mathcal{D}_{k}\right)}\right)\!.
\end{equation}
}

\textcolor{black}{$\widetilde{Ce^{t}}\left(v_{ij}\left(\mathcal{D}_{k}\right)\right)$
is the average rate of available SGLs by $v_{ij}\left(\mathcal{D}_{k}\right)$.
If there is no available SGL, $\widetilde{Ce^{t}}\left(v_{ij}\left(\mathcal{D}_{k}\right)\right)$=0.
}

\textcolor{black}{Since the satellites affect the state information
of each other when transmitting missions, to achieve efficient mission
scheduling of the whole network, we focus on a multi-agent network
scenario and adopt the joint state information (JSI) of multiple satellites
to make a decision. Specifically, in the BMS problem, we set each
satellite to exchange its local state information with its intra-domain
relays, and processes them. Further, each satellite constructs the
JSI of the intra-domain using processed state information of its own
and intra-domain relays, expressed as
\begin{equation}
\textrm{S}_{BMS}^{t}\left(v_{ij}\left(\mathcal{D}_{k}\right)\right)\!=\!\left[\widehat{s^{t}}\left({\scriptstyle v_{ij}\left(\mathcal{D}_{k}\right)}\right)\right]\cup\left[\widehat{s^{t}}\left({\scriptstyle m}\right)\right]_{m\in\mathfrak{\mathbb{\mathcal{R}}}\left(v_{ij}\left(\mathcal{D}_{k}\right),\mathcal{D}_{k}\right)}\!,
\end{equation}
where $\widehat{s^{t}}\left({\scriptstyle v_{ij}\left(\mathcal{D}_{k}\right)}\right)\!=\!\left(\!\widetilde{\mathcal{B}^{t}}\left({\scriptstyle v_{ij}\left(\mathcal{D}_{k}\right)}\right),\!\widetilde{E^{t}}\left({\scriptstyle v_{ij}\left(\mathcal{D}_{k}\right)}\right),\!\widetilde{Ce^{t}}\left({\scriptstyle v_{ij}\left(\mathcal{D}_{k}\right)}\right)\!,\right.$
$\left.\widehat{\mathcal{R}s^{t}}\left({\scriptstyle v_{ij}\left(\mathcal{D}_{k}\right)}\right)\right)$
and $\widehat{s^{t}}\left({\scriptstyle m}\right)\!=\!\left(\!\widetilde{\mathcal{B}^{t}\!}\left({\scriptstyle m}\right)\!,\!\widetilde{E^{t}\!}\left({\scriptstyle m}\right)\!,\!\widetilde{Ce^{t}\!}\left({\scriptstyle m}\right)\!,\!\widehat{\mathcal{R}s^{t}\!}\left({\scriptstyle m}\right)\!\right)$
are the processed state information. $\widehat{\mathcal{R}s^{t}}\left({\scriptstyle v_{ij}\left(\mathcal{D}_{k}\right)}\right)=\nicefrac{\widetilde{\mathcal{R}s^{t}}\left(v_{ij}\left(\mathcal{D}_{k}\right)\right)}{\widetilde{\mathcal{R}s^{t}}\left(v_{ij}\left(\mathcal{D}_{k}\right)\right)}$
represents the relative value of the average of the remaining survival
time slots, and if $\widetilde{\mathcal{R}s^{t}}\left(v_{ij}\left(\mathcal{D}_{k}\right)\right)=0$,
$\widehat{\mathcal{R}s^{t}}\left({\scriptstyle v_{ij}\left(\mathcal{D}_{k}\right)}\right)=0$.
Similarly, $\widehat{\mathcal{R}s^{t}}\left(m\right)=\nicefrac{\widetilde{\mathcal{R}s^{t}}\left(v_{ij}\left(\mathcal{D}_{k}\right)\right)}{\widetilde{\mathcal{R}s^{t}}\left(m\right)}$,
and if $\widetilde{\mathcal{R}s^{t}}\left(m\right)=0$, $\widehat{\mathcal{R}s^{t}}\left(m\right)=\widetilde{\mathcal{R}s^{t}}\left(v_{ij}\left(\mathcal{D}_{k}\right)\right)$.}

\begin{figure*}
\textcolor{black}{
\begin{equation}
\textrm{Pr}_{BMS}^{t}\left(v_{ij}\left(\mathcal{D}_{k}\right)\right)=\begin{cases}
\stackrel[{\scriptstyle g=1}]{{\scriptstyle N}}{\sum}\mathcal{X}^{t}\left({\scriptstyle a_{BMS}^{t}\left(v_{ij}\left(\mathcal{D}_{k}\right)\right),es_{g}}\right), & if\:a_{BMS}^{t}\left(v_{ij}\left(\mathcal{D}_{k}\right)\right)=v_{ij}\left(\mathcal{D}_{k}\right),\\
\stackrel[{\scriptstyle g=1}]{{\scriptstyle N}}{\sum}\widetilde{\mathcal{X}^{t}}\left({\scriptstyle a_{BMS}^{t}\left(v_{ij}\left(\mathcal{D}_{k}\right)\right),es_{g}}\right), & if\:a_{BMS}^{t}\left(v_{ij}\left(\mathcal{D}_{k}\right)\right)\neq v_{ij}\left(\mathcal{D}_{k}\right)\cap\widetilde{Ce^{t}}\left(v_{ij}\left(\mathcal{D}_{k}\right)\right)=0\\
 & \cap\,missions\,originate\,from\,v_{ij}\left(\mathcal{D}_{k}\right),\\
0, & else,
\end{cases}\label{eq:BMS_pro}
\end{equation}
}

\rule[0.5ex]{2.05\columnwidth}{1pt}
\end{figure*}

\textcolor{black}{The feasible action set of the BMS problem $\mathcal{\textrm{A}}_{BMS}^{t}\left(v_{ij}\left(\mathcal{D}_{k}\right)\right)$
corresponds to the index of optional intra-domain relays and the earth
station, which characterizes the satellites that can be selected to
perform mission transmission during mission scheduling, expressed
as $\mathcal{\textrm{A}}_{BMS}^{t}\left(v_{ij}\left(\mathcal{D}_{k}\right)\right)\!\!=\!\!\left\{ a_{BMS}^{t}\left(v_{ij}\left(\mathcal{D}_{k}\right)\right)\!\!=\!\!p\left|l^{t}\left(v_{ij}\left(\mathcal{D}_{k}\right),p\right)\!=\!1,\!p\!\in\!\left\{ \!\mathfrak{\mathbb{\mathcal{R}}\!}\left(v_{ij}\left(\mathcal{D}_{k}\right)\!,\!\mathcal{D}_{k}\right)\right.\right.\right.$
$\left.\left.\cup v_{ij}\left(\mathcal{D}_{k}\right)\right\} \right\} ,$
where $v_{ij}\left(\mathcal{D}_{k}\right)$ is used to indicate that
the missions are transmitted to the earth station, if $\exists l^{t}\left(v_{ij}\left(\mathcal{D}_{k}\right),es_{g}\right)\!=\!1$,
$l^{t}\left(v_{ij}\left(\mathcal{D}_{k}\right),v_{ij}\left(\mathcal{D}_{k}\right)\right)\!=\!1$,
otherwice, $l^{t}\left(v_{ij}\left(\mathcal{D}_{k}\right),v_{ij}\left(\mathcal{D}_{k}\right)\right)=0$.
$l^{t}\left(v_{ij}\left(\mathcal{D}_{k}\right),es_{g}\right)$ is
the connection relationship of SGL.}\textcolor{blue}{{} }\textcolor{black}{}

\textcolor{black}{For the reward of the BMS problem, we set it consists
of two parts, namely, 1) the profit value $\textrm{Pr}_{BMS}^{t}\left(v_{ij}\left(\mathcal{D}_{k}\right)\right)$:
the data volume of missions successfully transmitted to the earth
station, denoted as (\ref{eq:BMS_pro}), where $\widetilde{\mathcal{X}^{t}}\left({\scriptstyle a_{BMS}^{t}\left(v_{ij}\left(\mathcal{D}_{k}\right)\right),es_{g}}\right)$
is the data volume of missions from $v_{ij}\left(\mathcal{D}_{k}\right)$
successfully transmitted to the earth station via the transmission
mode 2);}\textcolor{blue}{{} }\textcolor{black}{2) the penalty value
$\textrm{Pe}_{BMS}^{t}\left(v_{ij}\left(\mathcal{D}_{k}\right)\right)$:
the data volume of missions failed to be received by the relay when
using the transmission mode 2), denoted as
\begin{equation}
\begin{array}{cc}
\!\!\textrm{Pe}_{BMS}^{t}\left({\scriptstyle {\textstyle v_{ij}\left(\mathcal{D}_{k}\right)}}\right)\!= & \!\!\!\!\!\!\!\!\!\!\mathcal{X}^{t}\left({\scriptstyle {\textstyle v_{ij}\left(\mathcal{D}_{k}\right),a_{BMS}^{t}\left(\!v_{ij}\left(\mathcal{D}_{k}\right)\!\right)}}\right)\\
 & \!\!\!\!\!\!\!\!\!\!-\widehat{\mathcal{X}^{t}}\left({\textstyle v_{ij}\left(\mathcal{D}_{k}\right),a_{BMS}^{t}\left(\!v_{ij}\left(\mathcal{D}_{k}\right)\!\right)}\right),
\end{array}\label{BMS_pe}
\end{equation}
then we define that (\ref{eq:BMS_pro}) minus (\ref{BMS_pe}) as the
reward, i.e., }\textcolor{blue}{}\textcolor{black}{
\begin{equation}
\textrm{R}_{BMS}^{t}\left({\scriptstyle {\textstyle {\scriptstyle v_{ij}\left(\mathcal{D}_{k}\right)}}}\right)=\textrm{Pr}_{BMS}^{t}\left({\scriptstyle v_{ij}\left(\mathcal{D}_{k}\right)}\right)-\textrm{Pe}_{BMS}^{t}\left({\scriptstyle v_{ij}\left(\mathcal{D}_{k}\right)}\right).\label{eq:reward}
\end{equation}
}

\textcolor{black}{To sum up, we can define that the BMS policy $\pi_{BMS}$
is a mapping from $\textrm{S}_{BMS}^{t}\left(v_{ij}\left(\mathcal{D}_{k}\right)\right)$
to $a_{BMS}^{t}\left(v_{ij}\left(\mathcal{D}_{k}\right)\right)$.
To measure the policy $\pi_{BMS}$, we define the state-value functions
to find an optimal policy that maximizes the long-term reward in the
intra-domain as follows:}

\textcolor{black}{
\begin{equation}
\textrm{BMS:}\:\max\;\enskip\mathbb{E}_{\pi,S}\left[\stackrel[{\scriptstyle t=1}]{{\scriptstyle T}}{\sum}\textrm{R}_{BMS}^{t}\left({\scriptstyle {\textstyle v_{ij}\left(\mathcal{D}_{k}\right)}}\right)\right],
\end{equation}
where $\mathbb{E}\left[\bullet\right]$ is the expectation function.}

\subsubsection{\textcolor{black}{TMS Problem }}

\textcolor{black}{In the TMS problem, we define it in a similar way
to the BMS problem.}\textcolor{blue}{{} }\textcolor{black}{Specifically,
each satellite constructs the JSI of the inter-domain using state
information of inter-domain relays and the average of the JSI of the
intra-domain, expressed as
\begin{equation}
\begin{array}{cc}
{\textstyle \textrm{S}_{TMS}^{t}\left({\scriptstyle v_{ij}\left(\mathcal{D}_{k}\right)}\right)=}\!\!\!\!\!\!\!\! & \left[\overline{\textrm{S}_{BMS}^{t}\left({\scriptstyle v_{ij}\left(\mathcal{D}_{k}\right)}\right)}\right]\cup\left[\widehat{s^{t}}\left({\scriptstyle m}\right)\right]_{m\in}\\
 & {\scriptscriptstyle \left\{ \mathfrak{\mathbb{\mathcal{R}}}\left(v_{ij}\left(\mathcal{D}_{k}\right),\mathcal{D}_{k'}\right)\left|k'\!\in\!\left\{ 1,2,\cdots,K\right\} ,k'\neq k\right.\right\} }
\end{array}
\end{equation}
where $\overline{\textrm{S}_{BMS}^{t}\left({\scriptstyle v_{ij}\left(\mathcal{D}_{k}\right)}\right)}$
is the average of the joint state information of the intra-domain,
denoted as
\begin{equation}
{\scriptstyle \overline{\textrm{S}_{BMS}^{t}\left({\scriptstyle v_{ij}\left(\mathcal{D}_{k}\right)}\right)}=\frac{\left[\widehat{s^{t}}\left({\scriptstyle v_{ij}\left(\mathcal{D}_{k}\right)}\right)\right]+\left[\widehat{s^{t}}\left({\scriptstyle m}\right)\right]_{m\in\mathfrak{\mathbb{\mathcal{R}}}\left(v_{ij}\left(\mathcal{D}_{k}\right),\mathcal{D}_{k}\right)}}{\left|\mathfrak{\mathbb{\mathcal{R}}}\left(v_{ij}\left(\mathcal{D}_{k}\right),\mathcal{D}_{k}\right)\right|+1}.}
\end{equation}
where $\left|\bullet\right|$ represents getting the number of elements
in a set.}

\begin{figure*}
\textcolor{black}{
\begin{equation}
\textrm{Pr}_{TMS}^{t}\left(v_{ij}\left(\mathcal{D}_{k}\right)\right)=\begin{cases}
\textrm{Pr}_{BMS}^{t}\left(v_{ij}\left(\mathcal{D}_{k}\right)\right), & if\:a_{TMS}^{t}\left(v_{ij}\left(\mathcal{D}_{k}\right)\right)=\mathcal{D}_{k},\\
\stackrel[{\scriptstyle g=1}]{{\scriptstyle N}}{\sum}\widetilde{\mathcal{X}^{t}}\left(a_{BMS}^{t}\left(v_{ij}\left(\mathcal{D}_{k'}\right)\right),es_{g}\right), & if\:a_{TMS}^{t}\left(v_{ij}\left(\mathcal{D}_{k}\right)\right)\neq\mathcal{D}_{k}\cap\,a_{BMS}^{t}\left(v_{ij}\left(\mathcal{D}_{k'}\right)\right)=v_{ij}\left(\mathcal{D}_{k'}\right)\\
 & \cap\,\widetilde{Ce^{t}}\left(v_{ij}\left(\mathcal{D}_{k}\right)\right)=0\cap\,missions\,originate\,from\,v_{ij}\left(\mathcal{D}_{k}\right),\\
0, & else,
\end{cases}\label{eq:TMS_pro}
\end{equation}
}

\rule[0.5ex]{2.05\columnwidth}{1pt}
\end{figure*}

\textcolor{black}{The feasible action set of the TMS problem $\mathcal{\textrm{A}}_{TMS}^{t}\left(v_{ij}\left(\mathcal{D}_{k}\right)\right)$
corresponds to the optional service domains in the satellite network,
which characterizes the domains that can be selected to perform mission
transmission during mission scheduling, expressed as  $\mathcal{\textrm{A}}_{TMS}^{t}\left(v_{ij}\left(\mathcal{D}_{k}\right)\right)\!\!=\!\!\left\{ \!a_{TMS}^{t}\!\left({\scriptstyle v_{ij}\left(\mathcal{D}_{k}\right)}\right)\!\!=\!\!\mathcal{D}_{k'}\!\left|\!\left\{ \!l^{t}\left(\!{\textstyle v_{ij}\!\left(\mathcal{D}_{k}\right)\!}{\textstyle ,\!p}\right)\!=\!1,\!p\!\in\!\mathfrak{\mathbb{\mathcal{R}}\!}\left(\!{\scriptstyle v_{ij}\left(\mathcal{D}_{k}\right)\text{,}\mathcal{D}_{k'}}\right)\!\right\} \!\neq\!\emptyset\!,\right.\right.$
$\left.{\scriptstyle k'\!\in\!\left\{ 1,2,\cdots,K\right\} ,k'\neq k}\right\} \cup\left\{ \!a_{TMS}^{t}\left({\scriptstyle v_{ij}\left(\mathcal{D}_{k}\right)}\right)\!\!=\!\!\mathcal{D}_{k}\left|\!\mathcal{\textrm{A}}_{BMS}^{t}\left({\scriptstyle v_{ij}\left(\mathcal{D}_{k}\right)}\right)\!\neq\!\emptyset\!\right.\right\} $.}

\textcolor{black}{Since the TMS policy is the domain of the transmission
mission, the data volume of the transmitted missions cannot be directly
obtained by executing this policy, i.e., the actual reward cannot
be directly obtained. However, after executing the TMS policy, CSs
will further execute the BMS policy to select one satellite for mission
transmission in the service domain determined by TMS. Therefore, the
reward of the TMS problem is obtained after executing the BMS policy
and also has two parts, 1) the profit value $\textrm{Pr}_{TMS}^{t}\left(v_{ij}\left(\mathcal{D}_{k}\right)\right)$:
the data volume of missions successfully transmitted to the earth
station by executing the TMS and BMS policies, denoted as (\ref{eq:TMS_pro}),
and 2) the penalty value $\textrm{Pe}_{TMS}^{t}\left(v_{ij}\left(\mathcal{D}_{k}\right)\right)$:
the total data volume of missions failed to be received by its relay
when using the transmission mode 2), denoted as
\begin{equation}
{\scriptstyle \textrm{Pe}_{TMS}^{t}\left(v_{ij}\left(\mathcal{D}_{k}\right)\right)}\!\!=\!\!\begin{cases}
{\scriptscriptstyle \textrm{Pe}_{BMS}^{t}\left(v_{ij}\left(\mathcal{D}_{k}\right)\right)}, & \!\!\!\!\!\!\!\!\!\!\!\!\!\!\!\!\!\!\!\!\!\!\!\!\!\!\!\!\!\!\!\!\!\!\!\!\!\!\!\!\!\!\!\!\!\!\!\!{\scriptscriptstyle if\:a_{TMS}^{t}\left(v_{ij}\left(\mathcal{D}_{k}\right)\right)=\mathcal{D}_{k},}\\
{\scriptscriptstyle \mathcal{X}^{t}\left(v_{ij}\left(\mathcal{D}_{k}\right),v_{ij}\left(\mathcal{D}_{k'}\right)\right)\!-\!\widehat{\mathcal{X}^{t}}\left(v_{ij}\left(\mathcal{D}_{k}\right),v_{ij}\left(\mathcal{D}_{k'}\right)\right)},\\
 & \!\!\!\!\!\!\!\!\!\!\!\!\!\!\!\!\!\!\!\!\!\!\!\!\!\!\!\!\!\!\!\!\!\!\!\!\!\!\!\!\!\!\!\!\!\!\!\!\!\!\!\!\!\!\!\!\!\!\!\!\!\!\!\!\!\!\!\!\!\!\!\!\!\!\!\!\!\!\!\!\!\!\!\!\!\!\!\!\!\!\!\!{\scriptscriptstyle if\:a_{TMS}^{t}\left(v_{ij}\left(\mathcal{D}_{k}\right)\right)\neq\mathcal{D}_{k},v_{ij}\left(\mathcal{D}_{k'}\right)\in\mathfrak{\mathbb{\mathcal{R}}}\left(v_{ij}\left(\mathcal{D}_{k}\right),\mathcal{D}_{k'}\right),}
\end{cases}
\end{equation}
then the reward is
\begin{equation}
\textrm{R}_{TMS}^{t}\left({\scriptstyle v_{ij}\left(\mathcal{D}_{k}\right)}\right)=\textrm{Pr}_{TMS}^{t}\left({\scriptstyle v_{ij}\left(\mathcal{D}_{k}\right)}\right)-\textrm{Pe}_{TMS}^{t}\left({\scriptstyle v_{ij}\left(\mathcal{D}_{k}\right)}\right).\label{eq:reward-1}
\end{equation}
}

\textcolor{black}{To sum up, we can define that the TMS policy $\pi_{TMS}$
is a mapping from $\textrm{S}_{TMS}^{t}\left(v_{ij}\left(\mathcal{D}_{k}\right)\right)$
to $a_{TMS}^{t}\left(v_{ij}\left(\mathcal{D}_{k}\right)\right)$.
Similarly, the state-value function of TMS is as follows:}

\textcolor{black}{
\begin{equation}
\textrm{TMS:}\:\max\;\enskip\mathbb{E}_{\pi,S}\left[\stackrel[{\scriptstyle t=1}]{{\scriptstyle T}}{\sum}\textrm{R}_{TMS}^{t}\left(v_{ij}\left(\mathcal{D}_{k}\right)\right)\right].
\end{equation}
}\vspace{-0.5em}

\begin{figure*}
\begin{equation}
\mathcal{L}_{BMS}\left({\scriptstyle \vartheta_{BMS}\left(v_{ij}\left(\mathcal{D}_{k}\right)\right)}\right)=-\frac{1}{\left|M_{BMS}\right|}\cdot\!\!\!\underset{{\scriptstyle t\in M_{BMS}}}{\sum}\!\!\!\log\left(\pi_{BMS}\left({\scriptstyle a_{BMS}^{t}\left(v_{ij}\left(\mathcal{D}_{k}\right)\right)\left|\textrm{S}_{BMS}^{t}\left(v_{ij}\left(\mathcal{D}_{k}\right)\right)\right.,\vartheta_{BMS}\left(v_{ij}\left(\mathcal{D}_{k}\right)\right)}\right)\right)\cdot W_{BMS}^{t}\left({\scriptstyle v_{ij}\left(\mathcal{D}_{k}\right)}\right).\label{eq:Loss_BMS_A}
\end{equation}

\begin{equation}
\mathcal{L}_{TMS}\left({\scriptstyle \vartheta_{TMS}\left(v_{ij}\left(\mathcal{D}_{k}\right)\right)}\right)=-\frac{1}{\left|M_{TMS}\right|}\cdot\!\!\!\underset{{\scriptstyle t\in M_{TMS}}}{\sum}\!\!\!\log\left(\pi_{TMS}\left({\scriptstyle a_{TMS}^{t}\left(v_{ij}\left(\mathcal{D}_{k}\right)\right)\left|\textrm{S}_{TMS}^{t}\left(v_{ij}\left(\mathcal{D}_{k}\right)\right)\right.,\vartheta_{TMS}\left(v_{ij}\left(\mathcal{D}_{k}\right)\right)}\right)\right)\cdot W_{TMS}^{t}\left({\scriptstyle v_{ij}\left(\mathcal{D}_{k}\right)}\right).\label{eq:Loss_TMS_A}
\end{equation}

\rule[0.5ex]{2.05\columnwidth}{1pt}

\begin{equation}
\mathcal{L}_{BMS}\left(\varpi_{BMS}\left(v_{ij}\left(\mathcal{D}_{k}\right)\right)\right)=\frac{1}{2\cdot\left|M_{BMS}\right|}\cdot\!\!\!\underset{{\scriptstyle t\in M_{BMS}}}{\sum}\!\!\!\left(\widehat{\textrm{R}_{BMS}^{t}}\left(v_{ij}\left(\mathcal{D}_{k}\right)\right)\!-\!\mathcal{V}_{BMS}\left(\textrm{S}_{BMS}^{t}\left(v_{ij}\left(\mathcal{D}_{k}\right)\right),\varpi_{BMS}\left(v_{ij}\left(\mathcal{D}_{k}\right)\right)\right)\right)^{2}.\label{eq:Loss_BMS_C}
\end{equation}

\begin{equation}
\mathcal{L}_{TMS}\left(\varpi_{TMS}\left(v_{ij}\left(\mathcal{D}_{k}\right)\right)\right)=\frac{1}{2\cdot\left|M_{TMS}\right|}\cdot\!\!\!\underset{{\scriptstyle t\in M_{TMS}}}{\sum}\!\!\!\left(\widehat{\textrm{R}_{TMS}^{t}}\left(v_{ij}\left(\mathcal{D}_{k}\right)\right)\!-\!\mathcal{V}_{TMS}\left(\textrm{S}_{TMS}^{t}\left(v_{ij}\left(\mathcal{D}_{k}\right)\right),\varpi_{TMS}\left(v_{ij}\left(\mathcal{D}_{k}\right)\right)\right)\right)^{2}.\label{eq:Loss_TMS_C}
\end{equation}

\rule[0.5ex]{2.05\columnwidth}{1pt}
\end{figure*}

\section{HICMS Algorithm Design\label{sec:HICMS-Algorithm-Design}\vspace{-0.5em}}

\textcolor{black}{To achieve higher adaptability and autonomy of
CMS and efficiently mitigate the impact of network scale, we develop
a HICMS algorithm to solve the HCMS problem. In this section, we first
elaborate on the framework of the proposed HICMS algorithm. Then,
we determine the local network environment of CSs and NCSs and acquire
the feasible action set of each satellite. Finally, we introduce the
setting and updating of the designed actor and critic networks.}\textcolor{blue}{{}
}\vspace{-0.5em}

\subsection{\textcolor{black}{HICMS Algorithm Framework}\vspace{-0.5em}}

\textcolor{black}{In the HICMS algorithm, we set each satellite in
the network environment as an agent to distributedly learn and execute
the CMS policy. Furthermore, since CSs undertake the intra-domain
and cross-domain mission's transmission and auxil}iary work, CSs will
fully implement the two stages of the HICMS algorithm, and NCSs only
need to implement the BMS stage.

\textcolor{black}{We take the CS as an example to introduce the process
of the HICMS algorithm to solve the HCMS problem as follows. First
of all, the CS determines the local network environment associated
with itself, which is used to obtain the joint state information and
the feasible action set. Then, the CS obtains the TMS policy using
the top-layer actor network according to the JSI of the inter-domain.
Execute the TMS policy to determine the service domain for mission
transmission, and according to the JSI of the intra-service domain,
apply the bottom-layer actor network to obtain the BMS policy. Execute
the BMS policy to complete the mission transmission to obtain the
actual reward and upload the reward obtained by the bottom layer to
the top layer. Finally, update the top- and bottom-layer critic and
actor networks to optimize the mission scheduling policy of the two
stages according to the obtained rewards of the BMS and TMS stages.
The above steps are performed iteratively until the whole network
obtains the converged and stable mission scheduling policy. }

\subsection{Determination\textcolor{black}{{} of the Local Network Environment\label{subsec:Determination-of-the}}}

\textcolor{black}{With the network scale increasing, each satellite
obtains the state information of all satellites in real-time is infeasible
in actual network scenarios \cite{Chu2020Multi}}\textcolor{blue}{.
}\textcolor{black}{Therefore, we set that each satellite obtains the
joint state information and the feasible action set from the local
network environment associated with itself. }

\textcolor{black}{The local network environment can be determined
according to the designed ISL and IDL establishment rules, i.e., the
local network environment of each satellite includes itself and its
relays.}\textcolor{blue}{{} }\textcolor{black}{Specifically, according
to the \textquotedbl one-satellite four-chain\textquotedbl{} mode,
by connecting to $v_{i\left(j-1\right)}\left(\mathcal{D}_{k}\right)$
and $v_{i\left(j+1\right)}\left(\mathcal{D}_{k}\right)$, $v_{ij}\left(\mathcal{D}_{k}\right)$
establishes the intra-plane ISL connections, and by connecting to
$v_{\left(i-1\right)j}\left(\mathcal{D}_{k}\right)$ and $v_{\left(i+1\right)j}\left(\mathcal{D}_{k}\right)$,
$v_{ij}\left(\mathcal{D}_{k}\right)$ establishes the inter-plane
ISL connections. Therefore, each satellite can select four satellites
that establish the intra- and inter-plane ISLs as the intra-domain
relays respectively. Since NCSs only have intra-domain relays, NCSs
can determine the local network environment according to the above
content. However, CSs also need to determine the inter-domain relays.
Before determining inter-domain relays, we first clear the number
of CSs and select the specific process of CSs. Due to the difference
in the number of satellites in each domain, we take the number of
satellites in the domain with the fewest satellites as the number
of CSs. Furthermore, we select satellites as CSs at equal intervals
in each domain until the number of CSs in this domain reaches the
upper limit. For example, if the domain with the fewest number of
satellites has 24 satellites, all satellites in this domain are CSs,
and if there are 48 satellites in other domains, one CS is selected
for every two satellites, and so on. Then, we set that each CS established
an IDL in each auxiliary domain according to the principle of proximity
and non-repetition to ensure that each CS can establish IDLs and the
number is consistent. Finally, each CS determines the satellite that
establishes IDLs with itself as inter-domain relays. To sum up, CSs
can determine the local network environment containing the intra-
and inter-domain relays.}

\subsection{\textcolor{black}{Acquisition of Feasible Action Set}}

\textcolor{black}{After determining the local network environment,
each satellite can acquire the feasible action set according to the
connection relationship of the SGLs, ISLs, and IDLs. However, due
to the high-speed orbit motion of satellites, the connection relationship
of the link is time-varying to cause the time-varying feasible action
set for each time slot. Therefore, we need to determine the connection
relationship of the SGLs, ISLs, and IDLs $\left(l^{t}\left(v_{ij}\left(\mathcal{D}_{k}\right),es_{g}\right)/l^{t}\left(v_{ij}\left(\mathcal{D}_{k}\right),m\right)\right)$
before selecting the policy to determine the feasible action set of
each satellite, where $l^{t}\left(v_{ij}\left(\mathcal{D}_{k}\right),m\right)$
can be obtained by the SRR and $l^{t}\left(v_{ij}\left(\mathcal{D}_{k}\right),es_{g}\right)$
can be obtained by Satellite Tool Kit (STK). }\textcolor{blue}{}

\textcolor{black}{For the BMS stage, each satellite needs to obtain
the connection relationship of SGLs and ISLs, and the set of the connection
relationship of the BMS stage is defined as $L_{BMS}^{t}\left(v_{ij}\left(\mathcal{D}_{k}\right)\right)=\left[l^{t}\left(v_{ij}\left(\mathcal{D}_{k}\right),m\right)\right]_{m\in\mathfrak{\mathbb{\mathcal{R}}}\left(v_{ij}\left(\mathcal{D}_{k}\right),\mathcal{D}_{k}\right)}\cup\left[l^{t}\left(v_{ij}\left(\mathcal{D}_{k}\right),es_{g}\right)\right]_{g\in\left\{ 1,2,\cdots,N\right\} }$.
The feasible action set of BMS stage $\textrm{A}_{BMS}^{t}\left(v_{ij}\left(\mathcal{D}_{k}\right)\right)$
can be acquired according to $L_{BMS}^{t}\left(v_{ij}\left(\mathcal{D}_{k}\right)\right)$.
For the TMS stage, CSs need to obtain the the connection relationship
of IDLs while obtaining the connection relationship of SGLs and ISLs.
The set of the connection relationship of the TMS stage is defined
as $L_{TMS}^{t}\left(v_{ij}\left(\mathcal{D}_{k}\right)\right)=L_{BMS}^{t}\left(v_{ij}\left(\mathcal{D}_{k}\right)\right)\cup\left[l^{t}\left(v_{ij}\left(\mathcal{D}_{k}\right),m\right)\right]_{m\in\left\{ \mathfrak{\mathbb{\mathcal{R}}}\left(v_{ij}\left(\mathcal{D}_{k}\right),\mathcal{D}_{k'}\right)\left|k'\!\in\!\left\{ 1,2,\cdots,K\right\} ,k'\neq k\right.\right\} }$.
The feasible action set of TMS stage $\textrm{A}_{TMS}^{t}\left(v_{ij}\left(\mathcal{D}_{k}\right)\right)$
can be acquired according to $L_{TMS}^{t}\left(v_{ij}\left(\mathcal{D}_{k}\right)\right)$.}

\begin{algorithm}
\caption{\textcolor{black}{\label{alg:Hierarchical-Intelligent-Cross-d}Hierarchical
Intelligent Cross-domain Mission Scheduling }Algorithm}

\begin{algorithmic}[1]

\REQUIRE{\small{}$\left|M_{BMS}\right|,\left|M_{TMS}\right|$, $\gamma$,
$T,episode$, $\alpha_{BMS}^{\vartheta},$ $\alpha_{BMS}^{\varpi},$
$\alpha_{TMS}^{\vartheta}\!,$ $\alpha_{TMS}^{\varpi}$.}{\small\par}

\ENSURE{\small{}${\scriptstyle \!\vartheta_{BMS}\!\left(\!{\scriptstyle v_{ij}\left(\mathcal{D}_{k}\right)}\!\right)},\!{\scriptstyle \vartheta_{TMS}\!\left(\!{\scriptstyle v_{ij}\left(\mathcal{D}_{k}\right)}\!\right),\varpi_{BMS}\!\left(\!{\scriptstyle v_{ij}\left(\mathcal{D}_{k}\right)}\!\right),}{\scriptstyle \varpi_{TMS}\!\left(\!{\scriptstyle v_{ij}\left(\mathcal{D}_{k}\right)\!}\right)}$}{\small\par}

\STATE{\footnotesize{} }{\small{}Initialize $\mathcal{Q}_{BMS}\!=\!0,\!\mathcal{Q}_{TMS}\!=\!0,\!M_{BMS}\!=\!\emptyset,\!M_{TMS}\!=\!\emptyset$.}{\small\par}

\STATE{\small{} Determine the local network environment of each satellite.\label{-Determine-the}}{\small\par}

\WHILE{{\small{}$episode>0$}}

\STATE {\small{}Initialize $\textrm{S}_{BMS}^{1}\left(v_{ij}\left(\mathcal{D}_{k}\right)\right),\textrm{S}_{TMS}^{1}\left(v_{ij}\left(\mathcal{D}_{k}\right)\right)$.\label{-Initialize-.}}{\small\par}

\FOR { {\small{}$t=\left\{ 1,2,\cdots,T\right\} $}}

\FOR {{\small{} each CS}{\footnotesize{} }}\label{--each}

\STATE {\small{}Acquire the feasible action set $\textrm{A}_{TMS}^{t}\left(v_{ij}\left(\mathcal{D}_{k}\right)\right)$.}{\small\par}

\STATE{\footnotesize{} }{\small{}Choose the TMS policy $a_{TMS}^{t}\left(v_{ij}\left(\mathcal{D}_{k}\right)\right)$.}{\small\par}

\ENDFOR

\FOR { {\small{}each satellite }}\label{--each-1}

\STATE {\small{}Acquire the feasible action set $\textrm{A}_{BMS}^{t}\left(v_{ij}\left(\mathcal{D}_{k}\right)\right)$.}{\small\par}

\STATE{\small{} Choose the BMS policy $a_{BMS}^{t}\left(v_{ij}\left(\mathcal{D}_{k}\right)\right)$.}{\small\par}

\ENDFOR

\STATE {\small{}CSs execute $a_{TMS}^{t}\left(v_{ij}\left(\mathcal{D}_{k}\right)\right)$
and $a_{BMS}^{t}\left(v_{ij}\left(\mathcal{D}_{k}\right)\right)$
to obtain $\mathcal{\textrm{R}}_{TMS}^{t}\left(v_{ij}\left(\mathcal{D}_{k}\right)\right)$
and $\mathcal{\textrm{R}}_{BMS}^{t}\left(v_{ij}\left(\mathcal{D}_{k}\right)\right)$,
and NCSs execute $a_{BMS}^{t}\left(v_{ij}\left(\mathcal{D}_{k}\right)\right)$
to obtain $\mathcal{\textrm{R}}_{BMS}^{t}\left(v_{ij}\left(\mathcal{D}_{k}\right)\right)$.\label{-CSs-execute}}{\small\par}

\STATE{\small{} Get $\textrm{S}_{BMS}^{t+1}\left(v_{ij}\left(\mathcal{D}_{k}\right)\right),\textrm{S}_{TMS}^{t+1}\left(v_{ij}\left(\mathcal{D}_{k}\right)\right)$.}{\small\par}

\STATE {\small{}${\scriptstyle {\textstyle M_{TMS}\!=\!M_{TMS}\cup}}$$\left(\textrm{S}_{TMS}^{t}\left(v_{ij}\left(\mathcal{D}_{k}\right)\right),a_{TMS}^{t}\left(v_{ij}\left(\mathcal{D}_{k}\right)\right),\right.$
${\textstyle \left.\mathcal{\textrm{R}}_{TMS}^{t}\left({\scriptstyle v_{ij}\left(\mathcal{D}_{k}\right)}\right),t,\textrm{S}_{TMS}^{t+1}\left({\scriptstyle v_{ij}\left(\mathcal{D}_{k}\right)}\right)\right)}$,
$M_{BMS}\!=\!M_{BMS}$ $\cup\left(\textrm{S}_{BMS}^{t}\!\left(\!{\scriptstyle {\textstyle v_{ij}\left(\mathcal{D}_{k}\right)}}\right)\!,\!a_{BMS}^{t}\!\left(\!{\scriptstyle {\textstyle v_{ij}\left(\mathcal{D}_{k}\right)}}\!\right),\!\textrm{R}_{BMS}^{t}\!\left(\!{\scriptstyle {\textstyle v_{ij}\left(\mathcal{D}_{k}\right)}}\!\right)\!,t,\right.$
$\left.\textrm{S}_{BMS}^{t+1}\left(v_{ij}\left(\mathcal{D}_{k}\right)\right)\right)$.\label{--,M}}{\small\par}

\STATE {\small{}$\textrm{S}_{BMS}^{t}\!\left(\!{\scriptstyle {\textstyle v_{ij}\left(\mathcal{D}_{k}\right)}}\right)=\textrm{S}_{BMS}^{t+1}\!\left(\!{\scriptstyle {\textstyle v_{ij}\left(\mathcal{D}_{k}\right)}}\right),$
$\textrm{S}_{TMS}^{t}\left(v_{ij}\left(\mathcal{D}_{k}\right)\right)=\textrm{S}_{TMS}^{t+1}\left(v_{ij}\left(\mathcal{D}_{k}\right)\right).$}{\small\par}

\STATE {\small{}$\mathcal{Q}_{BMS}=\mathcal{Q}_{BMS}+1,$ $\mathcal{Q}_{TMS}=\mathcal{Q}_{TMS}+1.$}{\small\par}

\IF{{\small{} $\mathcal{Q}_{BMS}=\left|M_{BMS}\right|$} }\label{-Q-BMS}

\FOR { {\small{}each satellite}{\footnotesize{} }}

\STATE {\small{}Calculate estimated state-value $\widehat{\textrm{R}_{BMS}^{t}}\left(v_{ij}\left(\mathcal{D}_{k}\right)\right)$.}{\small\par}

\STATE{\small{} Use loss functions to update $\vartheta_{BMS}\left(v_{ij}\left(\mathcal{D}_{k}\right)\right)$
and $\varpi_{BMS}\left(v_{ij}\left(\mathcal{D}_{k}\right)\right)$.}{\small\par}

\ENDFOR

\STATE {\small{}$\mathcal{Q}_{BMS}=0,M_{BMS}=\emptyset$.}{\small\par}

\ENDIF\label{BMS-endif}

\IF{{\small{} $\mathcal{Q}_{TMS}=\left|M_{TMS}\right|$} }

\FOR { {\small{}each CS} }

\STATE {\small{}Calculate estimated state-value $\widehat{\textrm{R}_{TMS}^{t}}\left(v_{ij}\left(\mathcal{D}_{k}\right)\right)$.}{\small\par}

\STATE {\small{}Use loss functions to update $\vartheta_{TMS}\left(v_{ij}\left(\mathcal{D}_{k}\right)\right)$
and $\varpi_{TMS}\left(v_{ij}\left(\mathcal{D}_{k}\right)\right)$.}{\small\par}

\ENDFOR

\STATE {\small{}$\mathcal{Q}_{TMS}=0,M_{TMS}=\emptyset$.}{\small\par}

\ENDIF\label{endif}

\ENDFOR

\STATE {\small{}$episode=episode-1$.}{\small\par}

\ENDWHILE

\end{algorithmic}
\end{algorithm}

\subsection{\textcolor{black}{Setting and Updating of the Actor and Critic Networks}}

\textcolor{black}{The multi-agent Advantage Actor-Critic (A2C) algorithm
framework is considered in this paper, and the actor and critic networks
are constructed using the deep neural network \cite{AddZheng2023}
to approximate the policy function and state-value function. Specifically,
the policy functions of the BMS and TMS stages are defined as $\pi_{BMS}\left(\textrm{S}_{BMS}^{t}\left(v_{ij}\left(\mathcal{D}_{k}\right)\right),\vartheta_{BMS}\left(v_{ij}\left(\mathcal{D}_{k}\right)\right)\right)$
and $\pi_{TMS}\left(\textrm{S}_{TMS}^{t}\left(v_{ij}\left(\mathcal{D}_{k}\right)\right),\vartheta_{TMS}\left(v_{ij}\left(\mathcal{D}_{k}\right)\right)\right)$,
respectively. The state-value functions of the BMS and TMS stages
are defined as $\mathcal{V}_{BMS}\left({\textstyle \textrm{S}_{BMS}^{t}\left(v_{ij}\left(\mathcal{D}_{k}\right)\right),\varpi_{BMS}\left(v_{ij}\left(\mathcal{D}_{k}\right)\right)}\right)$
and $\mathcal{V}_{TMS}\left(\textrm{S}_{TMS}^{t}\left(v_{ij}\left(\mathcal{D}_{k}\right)\right),\varpi_{TMS}\left(v_{ij}\left(\mathcal{D}_{k}\right)\right)\right)$,
respectively. $\vartheta_{BMS}\left(v_{ij}\left(\mathcal{D}_{k}\right)\right)$,
$\vartheta_{TMS}\left(v_{ij}\left(\mathcal{D}_{k}\right)\right)$,
$\varpi_{BMS}\left(v_{ij}\left(\mathcal{D}_{k}\right)\right)$, and
$\varpi_{TMS}\left(v_{ij}\left(\mathcal{D}_{k}\right)\right)$}\textcolor{blue}{{}
}\textcolor{black}{are the optimization parameters of the actor and
critic networks, including weight and bias.}

\textcolor{black}{For the actor and critic networks of the two stages,
the separate full connection (FC) layers are set as input layers for
processing the resource and mission states in JSIs of intra- and inter-domain,
i.e., the first layer includes 4 FC layers. Then, all the first layers'
output results are combined to input into a new FC layer. The ac}tivation
function of each FC layer is set to the ReLU. For the output layer,
the actor networks of the two stages are softmax for obtaining the
probability of each action, and the critic networks of the two stages
are linear. The actor and critic networks of the two stages are updated
by loss functions\textcolor{blue}{, }\textcolor{black}{where the loss
functions of actor networks can be expressed as} (\ref{eq:Loss_BMS_A})
and (\ref{eq:Loss_TMS_A}), and the loss functions of critic networks
can be expressed as (\ref{eq:Loss_BMS_C}) and (\ref{eq:Loss_TMS_C}).
In the (\ref{eq:Loss_BMS_A})-(\ref{eq:Loss_TMS_C}), $M_{BMS}$ and
$M_{TMS}$ are the minibatch of the BMS and TMS stages, respectively.
$\left|M_{BMS}\right|$\textcolor{black}{{} and $\left|M_{TMS}\right|$
are the minibatch size. $W_{BMS}^{t}\left({\scriptstyle v_{ij}\left(\mathcal{D}_{k}\right)}\right)=\widehat{\textrm{R}_{BMS}^{t}}\left(v_{ij}\left(\mathcal{D}_{k}\right)\right)-\mathcal{V}_{BMS}\left({\textstyle \textrm{S}_{BMS}^{t}\left(v_{ij}\left(\mathcal{D}_{k}\right)\right),\varpi_{BMS}^{-}\left(v_{ij}\left(\mathcal{D}_{k}\right)\right)}\right)$
is Temporal-Difference errors. $\widehat{\!\textrm{R}_{BMS}^{t}}\!\left(\!v_{ij}\left(\mathcal{D}_{k}\right)\!\right)\!=\!\textrm{R}_{BMS}^{t}\left({\scriptstyle {\textstyle v_{ij}\left(\mathcal{D}_{k}\right)}}\right)\!+\gamma\cdot\mathcal{V}_{BMS}\left(\textrm{S}_{BMS}^{t+1}\left(v_{ij}\left(\mathcal{D}_{k}\right)\right),\varpi_{BMS}^{-}\left(v_{ij}\left(\mathcal{D}_{k}\right)\right)\right)$
is the estimated state value, and $\gamma\in\left[0,1\right)$ is
the discount factor. $W_{TMS}^{t}\left({\scriptstyle v_{ij}\left(\mathcal{D}_{k}\right)}\right)$
and $\widehat{\textrm{R}_{TMS}^{t}}\left(v_{ij}\left(\mathcal{D}_{k}\right)\right)$
are similar to $W_{BMS}^{t}\left({\scriptstyle v_{ij}\left(\mathcal{D}_{k}\right)}\right)$
and $\widehat{\textrm{R}_{BMS}^{t}}\left(v_{ij}\left(\mathcal{D}_{k}\right)\right)$.}

\begin{table*}
\caption{Major simulation parameters.\label{tab:Major-simulation-parameters}}

\centering{}{\footnotesize{}}%
\begin{tabular}{|>{\centering}m{4cm}|>{\centering}m{1.5cm}|c|c|c|c|c|}
\hline 
{\footnotesize{}Parameters} & \textcolor{black}{\scriptsize{}$\mathcal{D}_{1}$} & \textcolor{black}{\scriptsize{}$\mathcal{D}_{2}$} & \textcolor{black}{\scriptsize{}$\mathcal{D}_{3}$} & \textcolor{black}{\scriptsize{}$\mathcal{D}_{4}$ (Add\_1)} & \textcolor{black}{\scriptsize{}$\mathcal{D}_{5}$ (Add\_2)} & \textcolor{black}{\scriptsize{}$\mathcal{D}_{6}$ (Add\_3)}\tabularnewline
\hline 
\hline 
{\footnotesize{}Number of satellites} & {\footnotesize{}66 ($6*11$)} & {\footnotesize{}48 ($8*6$)} & {\footnotesize{}24 ($3*8$)} & {\footnotesize{}24 ($6*4$)} & {\footnotesize{}60 ($6*10$)} & {\footnotesize{}48 ($8*6$)}\tabularnewline
\hline 
{\footnotesize{}Orbit height, inclination} & {\scriptsize{}780Km, 86.4${^\circ}$} & {\scriptsize{}1336Km, 66${^\circ}$} & {\scriptsize{}19100Km, 64.8${^\circ}$} & {\scriptsize{}20200Km, 55${^\circ}$} & {\scriptsize{}1070Km, 85${^\circ}$} & {\scriptsize{}1414Km, 52${^\circ}$}\tabularnewline
\hline 
{\footnotesize{}The type of mission} & {\footnotesize{}CM} & {\footnotesize{}OM} & {\footnotesize{}NM} & {\footnotesize{}NM} & {\footnotesize{}CM} & {\footnotesize{}OM}\tabularnewline
\hline 
{\footnotesize{}Total number of missions} & {\footnotesize{}10560} & {\footnotesize{}1824} & {\footnotesize{}4800} & {\footnotesize{}4800} & {\footnotesize{}7200} & {\footnotesize{}1824}\tabularnewline
\hline 
\textcolor{black}{\scriptsize{}$\rho\left(\mathcal{M}_{k}\right),\mathcal{R}s\left(\mathcal{M}_{k}\right)$} & \textcolor{black}{\footnotesize{}1Gbits, 18} & \textcolor{black}{\footnotesize{}3Gbits, 72} & \textcolor{black}{\footnotesize{}0.5Gbits, 6} & \textcolor{black}{\footnotesize{}0.5Gbits, 6} & \textcolor{black}{\footnotesize{}1Gbits, 18} & \textcolor{black}{\footnotesize{}3Gbits, 72}\tabularnewline
\hline 
{\footnotesize{}$\tau,T,P_{set},P_{sst},P_{sr}P_{o},P_{h},\eta$} & \multicolumn{6}{c|}{{\footnotesize{}100s, 216, 20}\textcolor{black}{\footnotesize{}W,
20W, 10W, 5W, 20W, }{\footnotesize{}$75\%$}}\tabularnewline
\hline 
{\footnotesize{}$\mathcal{B}_{max}$, $E_{max}$} & \multicolumn{6}{c|}{{\footnotesize{}NCSs: $\mathcal{B}_{max}$=60Gbits, $E_{max}$=100KJ;
CSs: $\mathcal{B}_{max}$=120Gbits, $E_{max}$=200KJ}}\tabularnewline
\hline 
\end{tabular}{\footnotesize\par}
\end{table*}

\textcolor{black}{For the actor and critic networks training, the
orthogonal initializer is adopted \cite{saxe2013exact} and the RMSprop
is used as the gradient optimizer. Furthermore, in light of the importance
of normalization for training the actor and critic networks, the JSIs
of intra- and inter-domain are normalized and clipped to $[0,2]$.
The reward is clipped to $[-2,2]$ \cite{Chu2020Multi}.}

The detailed pseudo-code of our proposed HICMS algorithm for solving
the HCMS problem is given in Algorithm \ref{alg:Hierarchical-Intelligent-Cross-d}
and explained next. \textcolor{black}{In the algorithm, each episode
consists of $T$ time slots. $\alpha_{BMS}^{\vartheta},$ $\alpha_{BMS}^{\varpi},$
$\alpha_{TMS}^{\vartheta},$ and $\alpha_{TMS}^{\varpi}$ are the
learning rates for the actor and critic networks of the BMS and TMS
stages. Each satellite first determines the local network environment
based on Section \ref{subsec:Determination-of-the} (line \ref{-Determine-the}).
During the training, the joint state information is reinitialized
in each episode (line \ref{-Initialize-.}). For the CSs, the TMS
and BMS stages are executed to obtain the rewards (lines \ref{--each}
to \ref{-CSs-execute}); for the NCSs, the BMS stage is executed to
obtain the reward (lines \ref{--each-1} to \ref{-CSs-execute}).
Then, each satellite collects the experience by the above process
until enough samples are collected for minibatch updating (line \ref{--,M}).
CSs update the parameters of the actor and critic networks in the
BMS and TMS stages (lines \ref{-Q-BMS} to \ref{endif}), and NCSs
update the parameters of the actor and critic networks in the BMS
stage (lines \ref{-Q-BMS} to \ref{BMS-endif}). This process is repeated
until $episode=0$.}

\section{Simulations And Discussions\label{sec:Simulation-Results-And}}

\textcolor{black}{In this section, we divide an extensive simulation
results into three parts to demonstrate the performance of the proposed
HICMS algorithm from different perspectives: 1) analyze the performance
improvement brought by CMS; 2) compare the performance of different
cross-domain algorithms; 3) compare the performance of algorithms
under different network scales. For the performance comparison, four
additional approaches as follows are considered:}
\begin{itemize}
\item \textcolor{black}{Independent domain mission scheduling (IDMS) \cite{Bao2023Towards}:
The algorithm is implemented in the framework of the BMS stage, and
during mission offloading, the resources of each domain are independent
of each other}\textcolor{blue}{.}\textcolor{black}{{} }
\item \textcolor{black}{No cooperative mission scheduling (NCMS): }In this
scheme\textcolor{black}{, each satellite only uses transmission mode
1) to offload missions.}
\item Intelligent cross-domain mission scheduling\textcolor{black}{{} (ICMS)
\cite{Bao2023Towards}: The implementation process of the algorithm
is similar to that of the BMS stage. The difference is that the feasible
action set includes inter-domain relays.}
\item \textcolor{black}{Blind Transmission Selection (BTS): In this scheme,
the CMS policy is randomly selected from the feasible action set with
the same probability, which means that it is a blind CMS policy without
any consideration of resource and mission states.}
\end{itemize}

\subsection{Simulation Configuration}

\textcolor{black}{Our simulations are conducted on a satellite network
scenario with three domains and ten earth stations (Section \ref{subsec:Performance-Analysis-of}
and \ref{subsec:Performance-Comparisons-of}). Furthermore, to verify
the effect of network scales on the performance of the algorithm,
we add three more domains to the simulation scenario (Section \ref{subsec:Performance-Comparisons-under}).
The main parameters used in the simulation are listed in Table \ref{tab:Major-simulation-parameters},
where the orbit parameters of each domain are set according to the
actual satellite systems}\footnote{\textcolor{black}{Both common and burst missions can exist in each
domain, and the total number of missions is not consistent with Table
\ref{tab:Major-simulation-parameters} only when the effect of different
mission numbers is investigated.}}\textcolor{black}{. The initial remaining survival time slot of burst
missions is set as $3$. According to the set simulation parameters
and \cite{Wu2019Research}, we use (\ref{eq:priority}) to calculate
the priority of common missions and get the result $\mathcal{P}\left(\mathcal{M}_{3}\right)>\mathcal{P}\left(\mathcal{M}_{1}\right)>\mathcal{P}\left(\mathcal{M}_{2}\right)$
and missions of the same type have the same priority. Furthermore,
the planning cycle is set for 6 hours, and all connection relationships
of SGLs from 15 Oct. 2022 04:00:00 to 15 Oct. 2022 10:00:00 are obtained
by STK. The transmission rates of ISL and IDL are distributed within
$[80,160]$Mbps and the transmission rate of SGLs is 60Mbp according
to \cite{GOLKAR2015230} and \cite{Fu2020Integrated}. For the training
of actor and critic networks, we set $\gamma=0.99$, $\alpha_{BMS}^{\varpi}=\alpha_{TMS}^{\varpi}=1e-4$,
$\alpha_{BMS}^{\vartheta}=\alpha_{TMS}^{\vartheta}=2.5e-4$, $\left|M_{BMS}\right|=\left|M_{TMS}\right|=72$,
and $episode=500$.}

\begin{figure}
\begin{centering}
\includegraphics[viewport=30bp 30bp 742bp 561bp,clip,scale=0.3]{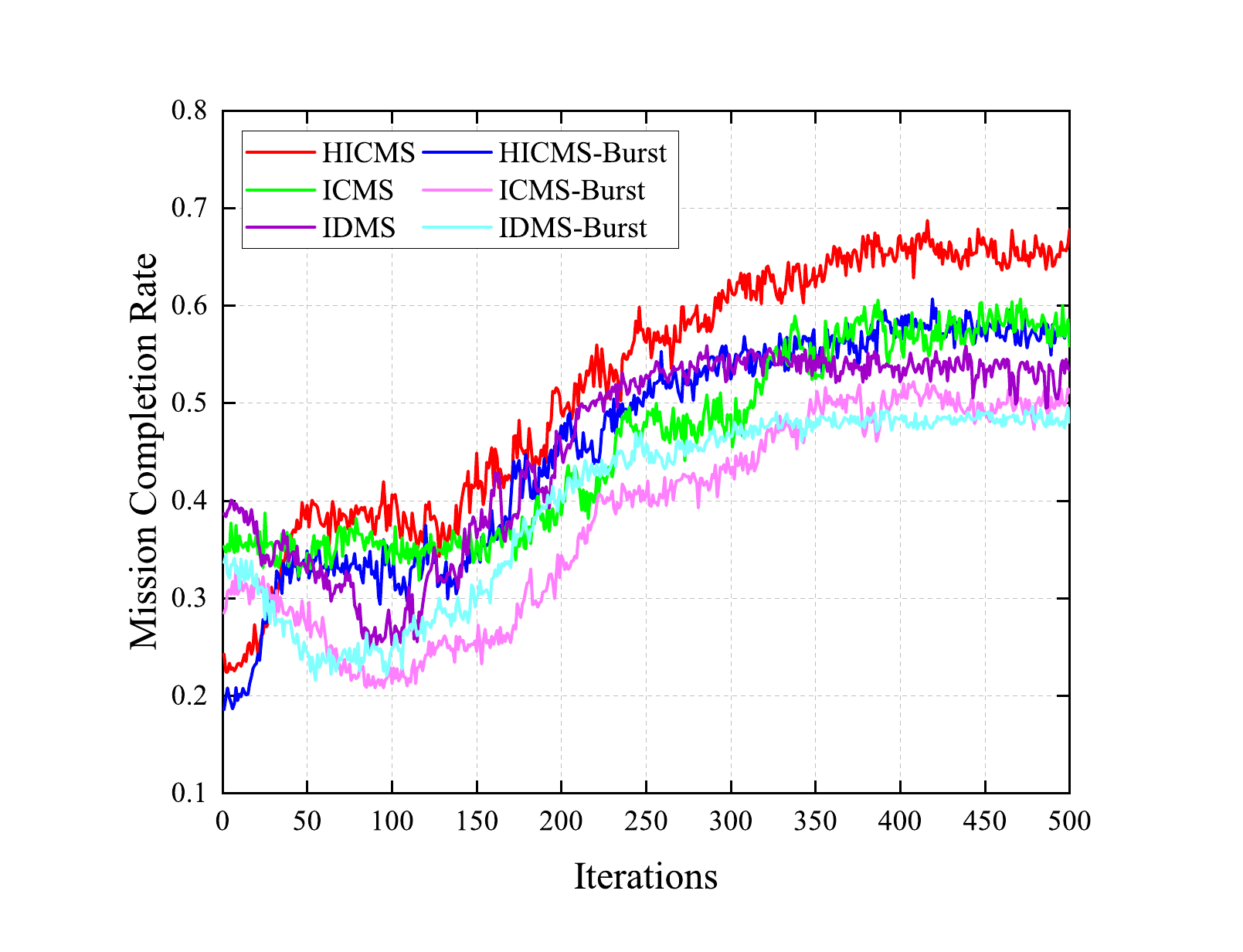}
\par\end{centering}
\caption{\textcolor{black}{\label{fig:Convergence-performance-of}Convergence
performance of the HICMS and existing algorithms.}}

\vspace{-1.0em}
\end{figure}

\subsection{Performance Evaluation}

\subsubsection{Performance Analysis of CMS\label{subsec:Performance-Analysis-of}}

\textcolor{black}{Figure \ref{fig:Convergence-performance-of} shows
the convergence performance of the HICMS and existing algorithms.
It can be seen that the mission completion rate (MCR) obtained by
the HICMS during the training process fluctuates and rises with the
optimization of the CMS policy and finally converges stably with small
fluctuations. Furthermore, whether burst missions are generated, the
HICMS has the best learning ability.}

\begin{figure}[t]
\begin{centering}
\includegraphics[viewport=30bp 30bp 742bp 561bp,clip,scale=0.3]{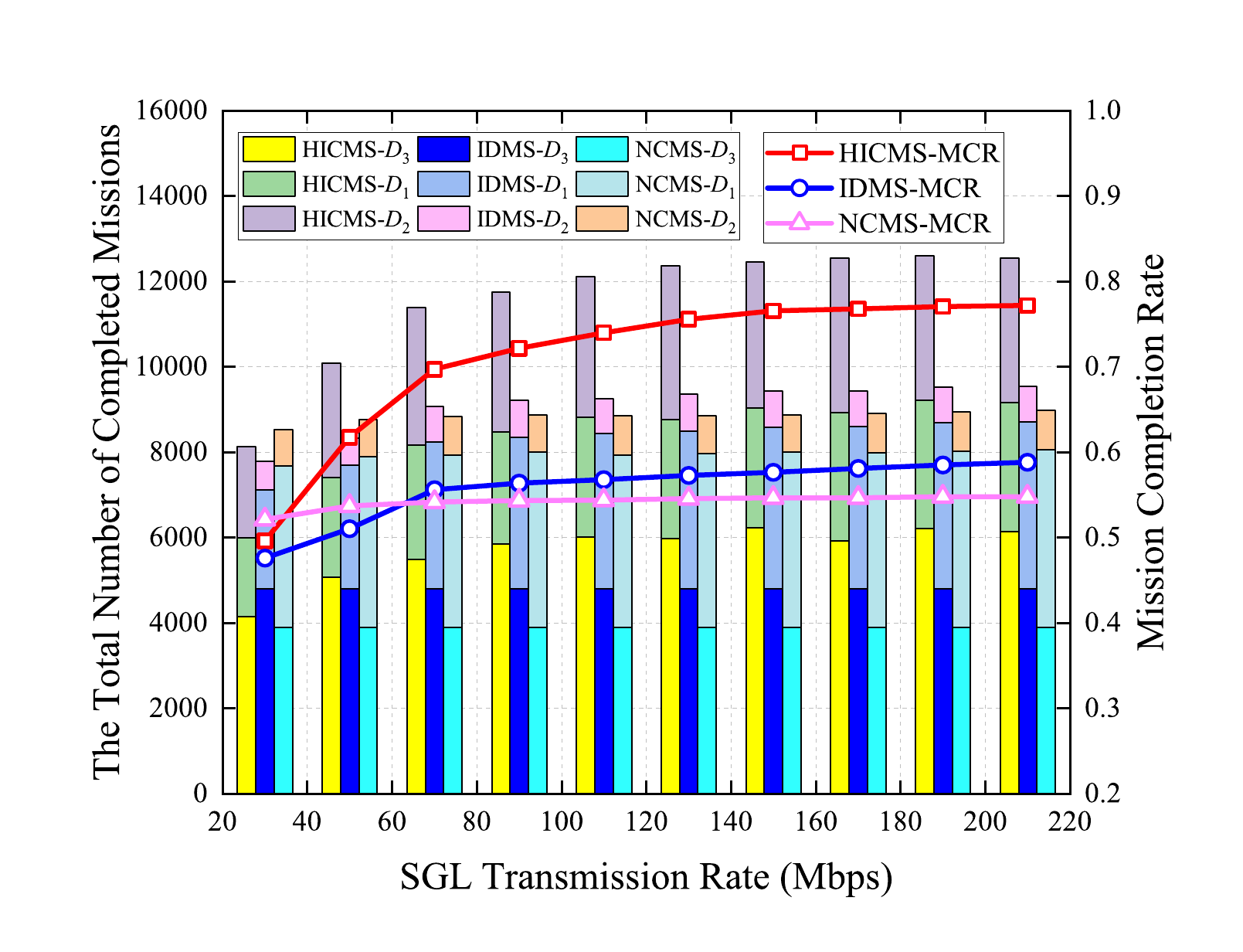}
\par\end{centering}
\begin{centering}
\caption{\textcolor{black}{\label{fig:The-total-number}The total number of
completed missions and MCR versus SGL transmission rate.}}
\par\end{centering}
\vspace{-1.0em}
\end{figure}

\begin{figure}
\begin{centering}
\includegraphics[viewport=30bp 30bp 742bp 561bp,clip,scale=0.3]{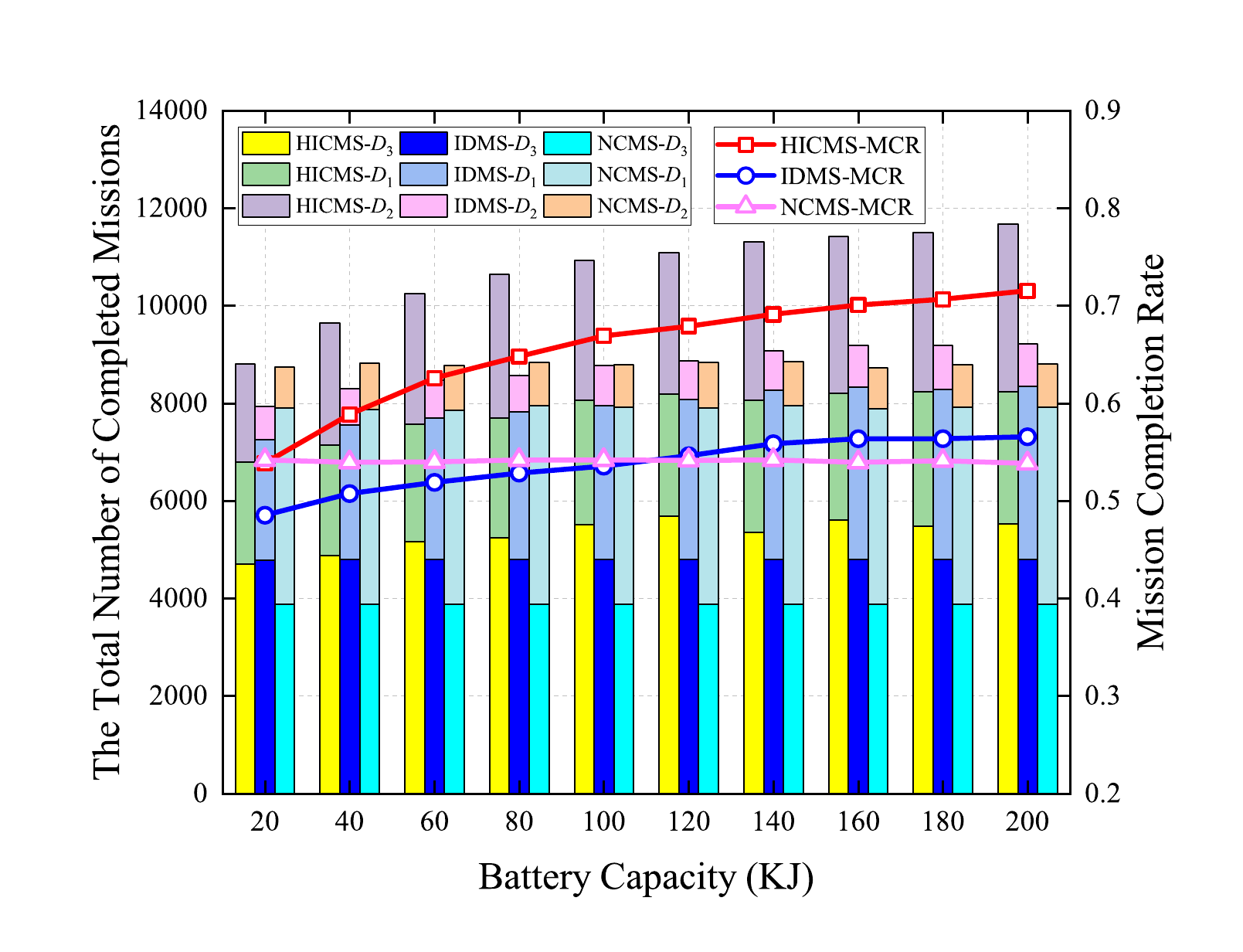}
\par\end{centering}
\begin{centering}
\caption{\textcolor{black}{\label{fig:The-total-number-1}The total number
of completed missions and MCR versus $E_{max}$.}}
\par\end{centering}
\vspace{-1.0em}
\end{figure}

\textcolor{black}{As shown in Figs. \ref{fig:The-total-number} and
\ref{fig:The-total-number-1}, we verify that the CMS leads to significant
performance gains under common missions. Specifically, HICMS achieves
higher MCR under different communication and energy resource configurations
through efficiently collaborating the inter-domain resources (For
example, the number of missions offloaded by $\mathcal{D}_{2}$ increases
significantly and exceeds the total number of missions collected $\mathcal{D}_{2}$.).
Further, with the increase of SGL transmission rate and battery capacity,
the gain of CMS increases gradually. The reasons are: 1) the increase
in transmission rate makes missions in each domain be offloaded in
time to reduce storage occupancy rate, which improves the possibility
of inter-domain resource collaboration; 2) CMS needs to consume more
energy and the improvement of battery capacity provides more energy
supply for CMS to improve mission offloading capability. Besides,
IDMS can obtain gains under sufficient communication and energy resources
through efficiently collaborating the intra-domain resources.}

\textcolor{black}{}

\begin{figure}
\begin{centering}
\includegraphics[viewport=30bp 30bp 742bp 561bp,clip,scale=0.3]{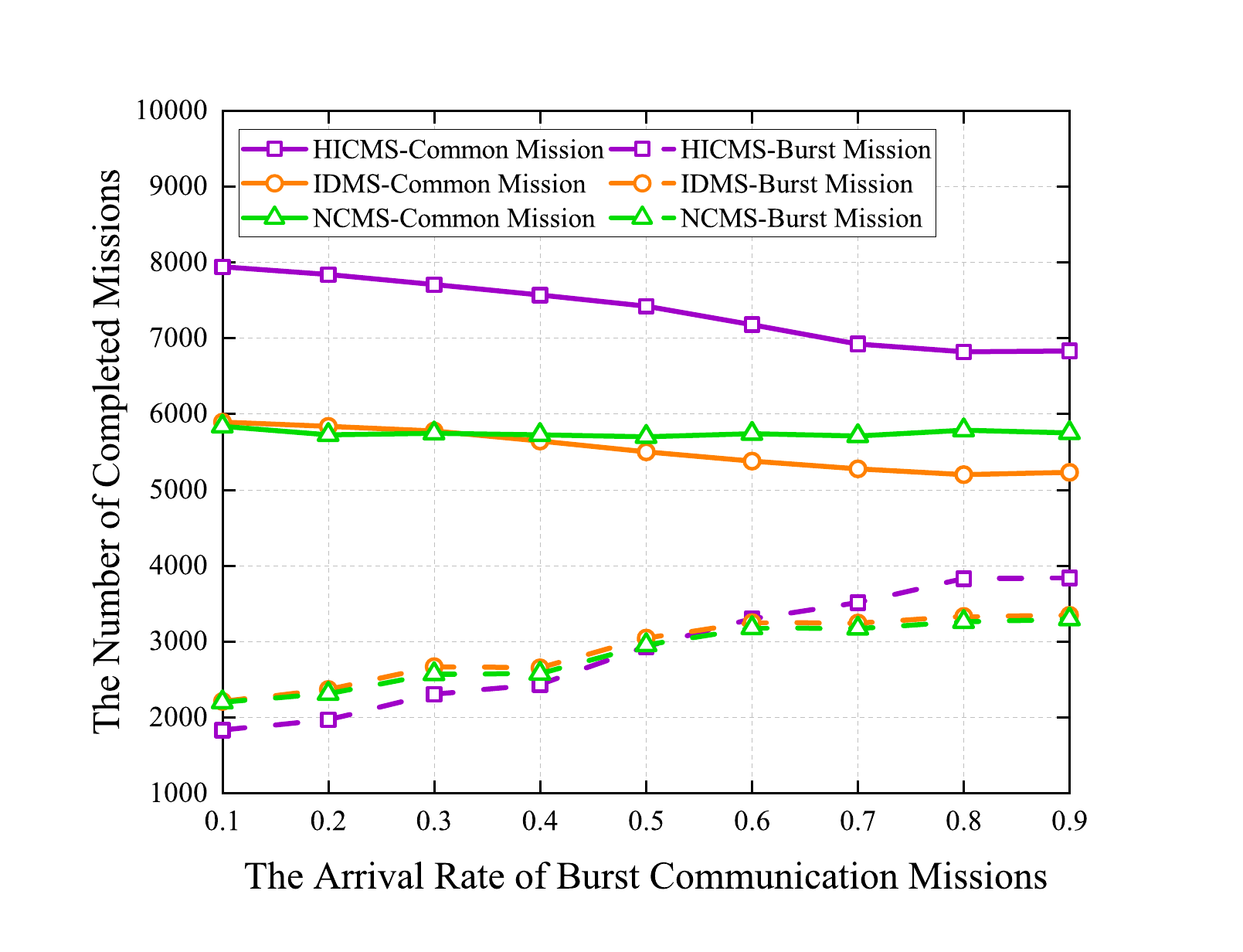}
\par\end{centering}
\caption{\textcolor{black}{\label{fig:The-number-of}The number of completed
missions versus the arrival rate of burst CMs.}}

\vspace{-1.0em}
\end{figure}

\begin{figure}[t]
\begin{centering}
\includegraphics[viewport=30bp 30bp 732bp 541bp,clip,scale=0.3]{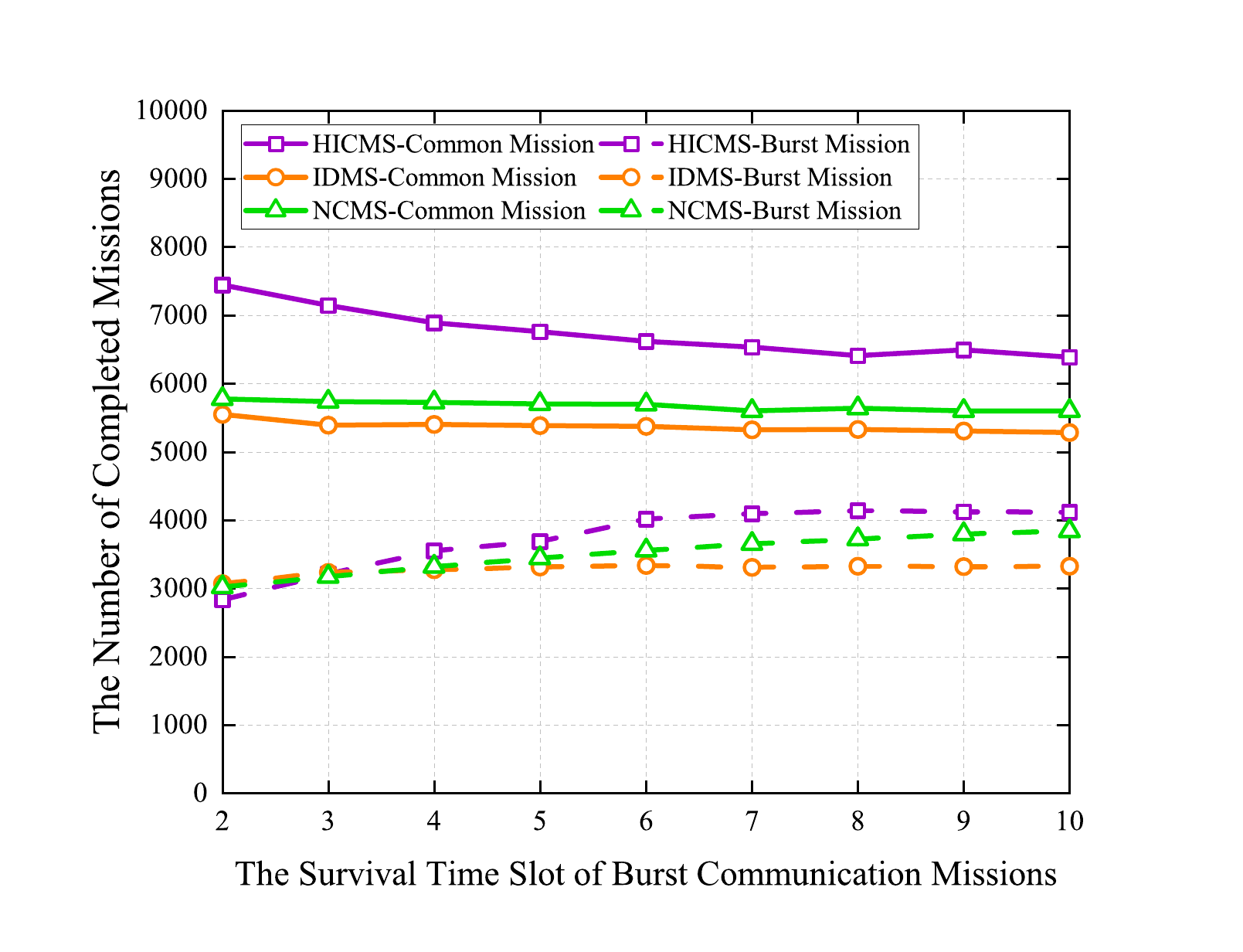}
\par\end{centering}
\begin{centering}
\caption{\textcolor{black}{\label{fig:The-number-of-1}The number of completed
missions versus the survival time slot of burst CMs.}}
\par\end{centering}
\vspace{-1.0em}
\end{figure}

\textcolor{black}{Figures \ref{fig:The-number-of} and \ref{fig:The-number-of-1}
show the number of completed common and burst missions under different
the arrival rates of burst CMs and under different survival time slots
of burst CMs to verify that the CMS can obtain better performance
under burst missions. In Fig. \ref{fig:The-number-of}, whether the
arrival rate of burst missions is large or small, the total number
of completed missions by the HICMS is the largest, and the total number
of completed missions by the three algorithms gradually increases.
In addition, with increasing the arrival rate of burst missions, the
number of completed burst missions by the HICMS gradually increases
and surpasses that of IDMS and NCMS. However, due to the highest priority
of burst missions, the increase in burst missions makes the offloading
of common missions lag, which reduces the number of completed common
missions. In Fig. \ref{fig:The-number-of-1}, with increasing the
survival time slot of burst CMs, the number of completed burst missions
by the three algorithms first increases and then stabilizes. Furthermore,
the total number of completed missions by the HICMS is the largest.
The reason is that the larger the survival time slot is, the more
opportunities the burst missions have to be successfully offloaded.
However, when the survival time slot is large, limited resources make
it impossible for the number of completed missions to grow continuously.
Furthermore, due to the long-term wait for burst missions to be transmitted,
the number of completed common missions is reduced.}

\begin{table*}
\caption{\label{tab:Algorithms-performance-compariso}\textcolor{black}{Different
algorithms' performance under different the number of domains.}}

\begin{centering}
{\scriptsize{}}%
\begin{tabular}{|>{\centering}m{1.5cm}|c|c|c|c|c|c|c|c|c|c|}
\hline 
\multirow{3}{1.5cm}{\centering{}{\scriptsize{}The number of domains}} & \multicolumn{10}{c|}{{\scriptsize{}The total number of completed missions}}\tabularnewline
\cline{2-11} \cline{3-11} \cline{4-11} \cline{5-11} \cline{6-11} \cline{7-11} \cline{8-11} \cline{9-11} \cline{10-11} \cline{11-11} 
 & \multicolumn{5}{c|}{{\scriptsize{}Common mission (Mode 1)}} & \multicolumn{5}{c|}{{\scriptsize{}Common and burst missions (Mode 2)}}\tabularnewline
\cline{2-11} \cline{3-11} \cline{4-11} \cline{5-11} \cline{6-11} \cline{7-11} \cline{8-11} \cline{9-11} \cline{10-11} \cline{11-11} 
 & {\scriptsize{}HICMS} & {\scriptsize{}IDMS} & {\scriptsize{}NCMS} & {\scriptsize{}ICMS} & {\scriptsize{}BTS} & {\scriptsize{}HICMS} & {\scriptsize{}IDMS} & {\scriptsize{}NCMS} & {\scriptsize{}ICMS} & {\scriptsize{}BTS}\tabularnewline
\hline 
\hline 
{\scriptsize{}2 domains} & \textbf{\scriptsize{}5395} & {\scriptsize{}3977} & {\scriptsize{}4954} & {\scriptsize{}5370} & {\scriptsize{}2462} & \textbf{\scriptsize{}4838}{\scriptsize{} }{\tiny{}(}\textbf{\tiny{}4002}{\tiny{}/836)} & {\scriptsize{}3591}{\tiny{} (2844/747)} & {\scriptsize{}4396 }{\tiny{}(3577/819)} & {\scriptsize{}4426 }{\tiny{}(3464/}\textbf{\tiny{}962}{\tiny{})} & {\scriptsize{}2405}{\tiny{} (2078/327)}\tabularnewline
\hline 
{\scriptsize{}3 domains} & \textbf{\scriptsize{}10927} & {\scriptsize{}8718} & {\scriptsize{}8794} & {\scriptsize{}9167} & {\scriptsize{}6006} & \textbf{\scriptsize{}10367}{\scriptsize{} }{\tiny{}(}\textbf{\tiny{}7150}{\tiny{}/}\textbf{\tiny{}3217}{\tiny{})} & {\scriptsize{}8294 }{\tiny{}(5684/2610)} & {\scriptsize{}8150 }{\tiny{}(5924/2226)} & {\scriptsize{}9114 }{\tiny{}(5961/3153)} & {\scriptsize{}5326 }{\tiny{}(3965/1361)}\tabularnewline
\hline 
{\scriptsize{}4 domains} & \textbf{\scriptsize{}15398} & {\scriptsize{}13587} & {\scriptsize{}12700} & {\scriptsize{}14497} & {\scriptsize{}10161} & \textbf{\scriptsize{}14229}{\tiny{} (}\textbf{\tiny{}10557}{\tiny{}/3672)} & {\scriptsize{}13125 }{\tiny{}(8689/}\textbf{\tiny{}4436}{\tiny{})} & {\scriptsize{}11818 }{\tiny{}(8161/3657)} & {\scriptsize{}13714}{\tiny{} (9436/4278)} & {\scriptsize{}8698}{\tiny{} (6391/2307)}\tabularnewline
\hline 
{\scriptsize{}5 domains} & \textbf{\scriptsize{}20566} & {\scriptsize{}17338} & {\scriptsize{}15973} & {\scriptsize{}17040} & {\scriptsize{}11737} & \textbf{\scriptsize{}17070}{\scriptsize{} }{\tiny{}(}\textbf{\tiny{}12701}{\tiny{}/4377)} & {\scriptsize{}15124 }{\tiny{}(10110/}\textbf{\tiny{}5014}{\tiny{})} & {\scriptsize{}14101}{\tiny{} (9746/4355)} & {\scriptsize{}15586 }{\tiny{}(11182/4404)} & {\scriptsize{}9953 }{\tiny{}(7476/2477)}\tabularnewline
\hline 
{\scriptsize{}6 domains} & \textbf{\scriptsize{}21071} & {\scriptsize{}18258} & {\scriptsize{}16824} & {\scriptsize{}18120} & {\scriptsize{}12583} & \textbf{\scriptsize{}18347}{\scriptsize{} }{\tiny{}(}\textbf{\tiny{}13704}{\tiny{}/4643)} & {\scriptsize{}16641 }{\tiny{}(11576/}\textbf{\tiny{}5065}{\tiny{})} & {\scriptsize{}15300}{\tiny{} (10690/4610)} & {\scriptsize{}15607 }{\tiny{}(10953/4654)} & {\scriptsize{}11012 }{\tiny{}(8367/2645)}\tabularnewline
\hline 
\end{tabular}{\scriptsize\par}
\par\end{centering}
\vspace{-0.5em}
\end{table*}

\begin{figure*}[t]
\begin{centering}
\begin{minipage}[t]{0.32\linewidth}%
\includegraphics[viewport=10bp 30bp 742bp 561bp,clip,scale=0.25]{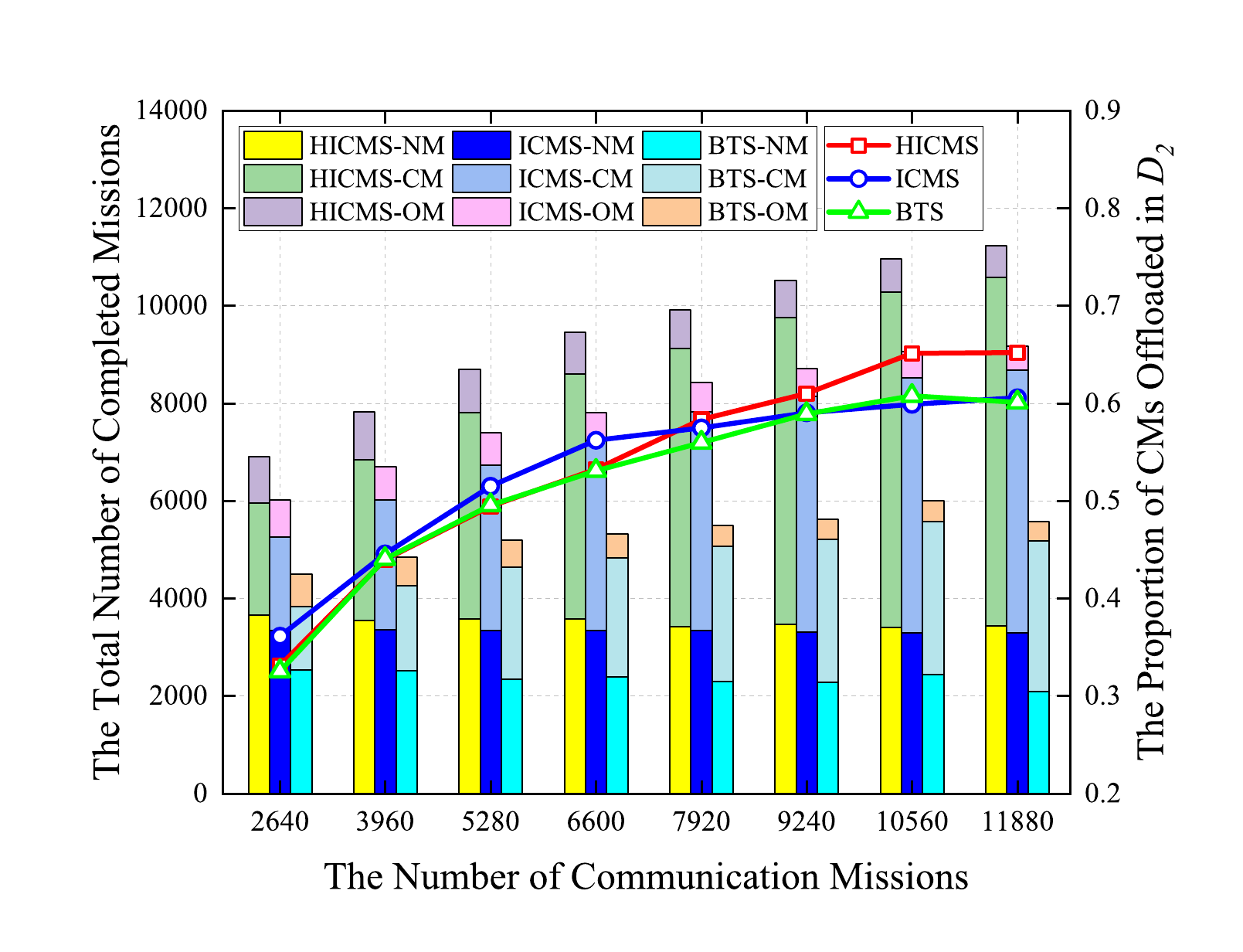}
\begin{center}
\caption{\textcolor{black}{\label{fig:The-number-of-3}The total number of
completed missions and the proportion of CMs offloaded in $\mathcal{D}_{2}$
under different the number of CMs.}}
\par\end{center}%
\end{minipage}\hfill{}%
\begin{minipage}[t]{0.32\linewidth}%
\includegraphics[viewport=20bp 30bp 732bp 541bp,clip,scale=0.25]{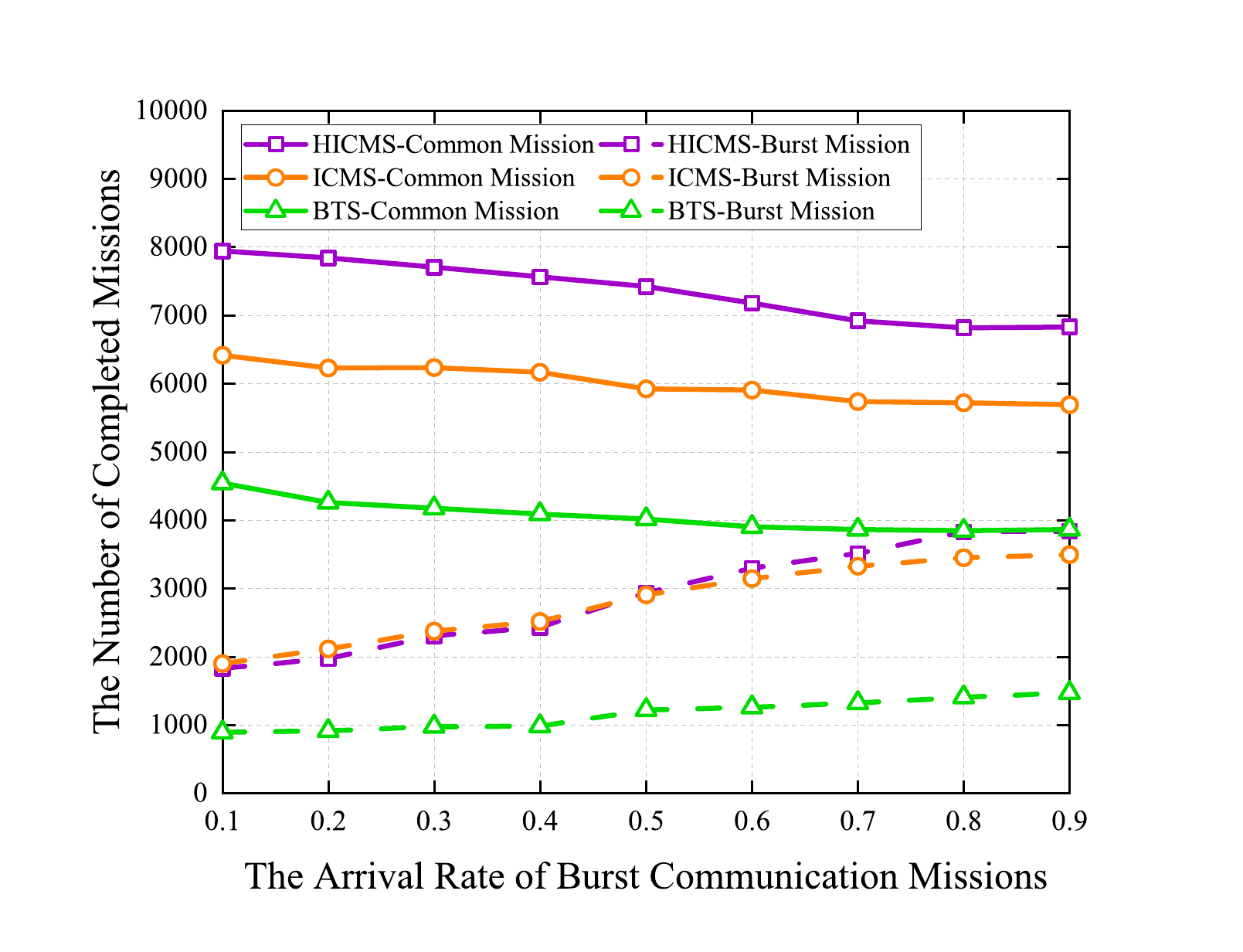}
\begin{center}
\caption{\textcolor{black}{\label{fig:The-number-of-4}The number of completed
missions versus the arrival rate of burst CMs.}}
\par\end{center}%
\end{minipage}\hfill{}%
\begin{minipage}[t]{0.32\linewidth}%
\includegraphics[viewport=30bp 30bp 732bp 541bp,clip,scale=0.25]{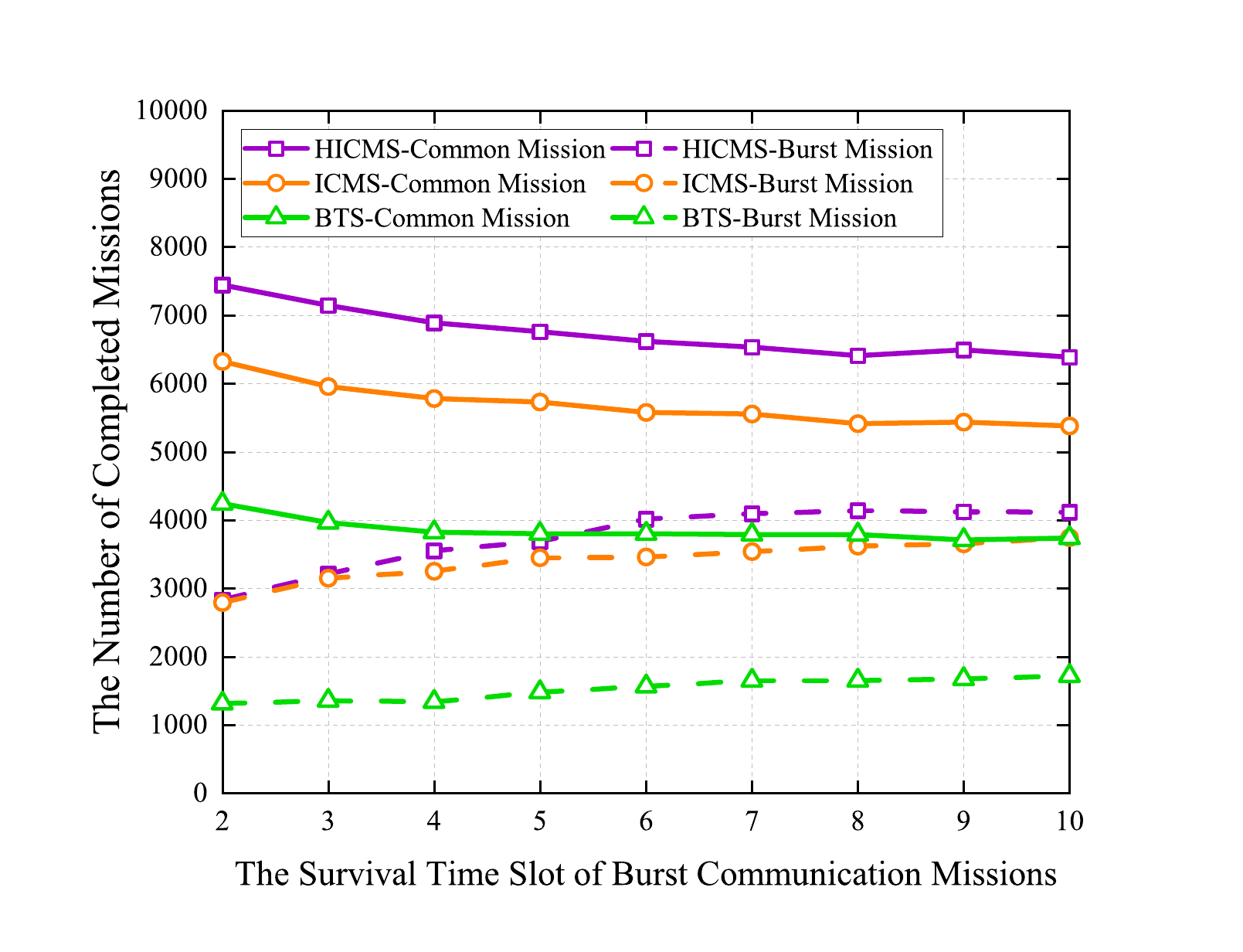}
\begin{center}
\caption{\textcolor{black}{\label{fig:The-number-of-5}The number of completed
missions versus the survival time slot of burst CMs.}}
\par\end{center}%
\end{minipage}
\par\end{centering}
\vspace{-0.5em}
\end{figure*}

\subsubsection{Performance Comparisons of Different Cross-domain Algorithms\label{subsec:Performance-Comparisons-of}}

\textcolor{black}{Figure \ref{fig:The-number-of-3} shows the total
number of completed missions and the proportion of CMs offloaded in
$\mathcal{D}_{2}$ under different the number of CMs. It can be seen
that with increasing the number of CMs, the performance of the HICMS
is better than the ICMS and BTS, and the total number of completed
missions increases gradually. Furthermore, for the three algorithms,
the increase of CMs does not affect the offloading of NMs, because
the priority of NMs is higher than that of CMs. However, the number
of offloaded OMs decreased gradually, which indicates that the significant
increase in the number of offloaded CMs is assisted by $\mathcal{D}_{2}$.
The proportion of CMs offloaded in $\mathcal{D}_{2}$ also effectively
proves this conclusion. It can be observed from the curve that the
proportion of CMs in the missions offloaded by $\mathcal{D}_{2}$
increases gradually, and the inter-domain collaboration performance
of the HICMS is better. Besides, since ICMS directly selects intra-domain/inter-domain
relays for cooperative mission scheduling and IDLs do not exist in
every time slot, inter-domain collaboration performance and the number
of completed missions are poorer than HICMS.}

\textcolor{black}{As shown in Figs. \ref{fig:The-number-of-4} and
\ref{fig:The-number-of-5}, whether common missions or burst missions,
HICMS obtains the best performance, and the total number of completed
missions gradually increases. In Fig. \ref{fig:The-number-of-4},
when the arrival rate of burst missions is small, the number of completed
burst missions of HICMS and ICMS is almost equal, and with the arrival
rate of burst missions increasing, the performance of the HICMS outperforms
that of the ICMS. The reason is that the inter-domain collaboration
of ICMS is poor, and burst missions cannot be completed through efficient
inter-domain resource collaboration when the intra-domain resources
cannot be supplied in time. Besides, since BTS cannot effectively
select the CMS policy based on resources and mission states, the total
number of completed missions hardly increases. In Fig. \ref{fig:The-number-of-5},
similarly, due to the difficulty in guaranteeing the effective offloading
of burst missions, with increasing the survival time slot of burst
CMs, the number of completed burst missions in ICMS increases slightly
and soon stabilizes. The total number of completed missions remains
almost unchanged for BTS.}

\subsubsection{\textcolor{black}{Performance Comparisons under Different Network
Scales\label{subsec:Performance-Comparisons-under}}}

\textcolor{black}{Table \ref{tab:Algorithms-performance-compariso}
shows the different algorithms' performance comparison under different
the number of domains. It can be observed that in Modes 1 and 2, the
total number of completed missions by HICMS is the highest under different
the number of domains. Furthermore, with increasing the number of
domains, the total number of completed missions by the IDMS is similar
to that of ICMS. The reason is that due to the poor inter-domain collaboration
of the ICMS, the more domains, the harder it is to learn a better
CMS policy to gradually loss the advantages of the CMS. In Mode 2,
the number of completed common missions by the HICMS is always the
highest, and the number of completed burst missions is not much different
from the highest one. IDMS has a good performance for burst missions,
which may be because burst missions with a higher priority can be
directly offloaded in the domain to avoid the reduction of the remaining
survival time slots. Besides, since the remaining survival time slots
of burst missions are the smallest, under the same total number of
missions, the number of completed missions in Mode 2 is less than
in Mode 1.}

\textcolor{black}{Furthermore, it should be noted that due to limited
transceivers of satellites, when the number of domains is greater
than 3, we set that each domain selects only two domains as auxiliary
domains and each CS establishes IDLs in the two auxiliary domains.
Therefore, for the same number of domains, the MCR obtained by selecting
different IDLs is also different, but the performance of HICMS is
still higher than that of the other algorithms.}

\vspace{-0.5em}

\section{Conclusion\label{sec:Conclusion}}

\textcolor{black}{In this paper, we investigate the CMS problem for
satellite networks to efficiently collaborate multi-domain resources.
First, we accurately characterize the communication resource state
of inter-satellite, and systematically characterize the differentiation
of mission demands. Hereafter, we convert the CMS problem to the HCMS
problem and develop the HICMS algorithm to solve it. The proposed
algorithm can dynamically adjust and efficiently match the CMS policy
to efficiently collaborate intra- and inter-domain resources to improve
mission completion performance. Simulation results demonstrate that
the proposed algorithm outperforms the independent domains and existing
CMS algorithms and can still guarantee performance in case the network
scales increase.}

\bibliographystyle{IEEEtran}
\bibliography{reference}

\end{document}